\documentclass[]{aa}
\usepackage{graphicx}

\def\0{\phantom0}
\def\Tastrom{\Delta t_{\rm astrom}}
\def\Tobs{\Delta t_{\rm obs}}
\def\Tpred{\Delta t_{\rm pred}}
\def\figskip{\vskip-\bigskipamount}
\def\zl{z_{\rm lens}}

\def\<#1>{[\hbox{#1}]}
\let\savecr\\ \def\\{\\noalign{\smallskip}}

\title{COSMOGRAIL: the COSmological MOnitoring of \\
   \vspace*{1mm} GRAvItational Lenses IV.}

\subtitle{Models of prospective time-delay lenses}

\begin{document}

\author{P. Saha \inst{1,2}
        \and
        F. Courbin \inst{3}
        \and
        D. Sluse \inst{3}
        \and
        S. Dye \inst{4}
        \and
        G. Meylan \inst{3}}

\institute{Astronomy Unit, 
           Queen Mary and Westfield College,
           University of London,
           London E1~4NS, UK.
           \and
           Instit\"ut f\"ur Theoretische Physik,
           Universit\"at Z\"urich,
           Winterthurerstr 190,
           CH-8057 Z\"urich, Switzerland.
           \and
           Laboratoire d'Astrophysique,
           Ecole Polytechnique F\'ed\'erale de Lausanne,
           Observatoire, CH-1290 Sauverny, Switzerland.
           \and
           School of Physics and Astronomy, Cardiff University, 
           5 The Parade, Cardiff, CF24 3YB, UK.
          }

\date{}

\abstract{}
{To predict time delays for a sample of gravitationally lensed quasars 
and to evaluate the accuracy that can be realistically 
achieved on the value of $H_0$.}
{We consider 14 lensed quasars that are candidates for
time-delay monitoring and model them in detail using pixelized lens
models. For each system, we provide a mass map, arrival-time surface
and the distribution of predicted time-delays in a concordance
cosmology, assuming $H_0^{-1}=14$~Gyr ($H_0=70$ in local units).
Based on the predicted time-delays and on the observational
circumstances, we rate each lens as `excellent' or `good' or
`unpromising' for time-delay monitoring.  Finally, we analyze
simulated time delays for the 11 lens rated excellent or good, and
show that $H_0$ can be recovered to a precision of 5\%.}
{In combination with COSMOGRAIL paper~I on the temporal 
sampling of lensed quasar light curves, the present
work will help design monitoring campaigns of lensed quasars.}
{}


\maketitle

\section{Introduction}

Gravitational lensing of distant quasars is one of many possible
routes to $H_0$. It has unique advantages. First, lensing depends on
well-understood physics: gravitation.  Second, time-delay observations
require modest resources, hence low-demand telescopes can make a
significant contribution.  Third, the galaxy models used to convert
time-delays into $H_0$ have made considerable progress in the past
decade.  As a result, time-delay measurements have become an
increasingly active research topic.\footnote{According to ADS, the
original paper by \cite{refsdal64} pointing out the connection between
gravitational lensing time-delays and $H_0$ was cited on-average once
every two years through the 1960s and 70s, whereas nowadays it is
cited about once every two weeks.}  So far there are 12 secure
time-delays (Table~\ref{tablobs}), of which 10 yield an estimate of
the Hubble constant --- the lens identification in PKS~1830--211
remains controversial (\cite{cou02}, \cite{winn02}), and the lensing
galaxy HE~0435--122 may be anomalous (Kochanek 2005).

The dominant uncertainty in measuring $H_0$ from lensing is the
non-uniqueness of lens mass profiles that can reproduce the
observables.  Before the non-uniqueness of mass-models was widely
appreciated, researchers would usually fit a single family of mass
models to data, leading to over-optimistic error bars.  Experimenting
with different kinds of mass model for the same data pointed to much
larger uncertainties (Schechter et al.\ 1997,
\cite{sw97,bernstein99,keeton00}).  More recently, procedures
involving sampling an ensemble of models according to some prior are
being preferred, in order to derive a more useful picture of the
uncertainties (\cite{ws00,sw04,oguri04a}, Jakobsson et al.\ 2005).  A
fair summary of current $H_0$ results from lensing is that the
error-bars are competitive on the young-Universe (i.e., high $H_0$)
side but the old-Universe side needs improvement.

Clearly, to reach the 5--10\% accuracy claimed by some other
techniques (e.g., \cite{freedman01,spergel03}), more time-delays are
needed.  But to run monitoring campaigns efficiently, it is important
to have preliminary estimates for time-delays --- witness the tenfold
range in the known values in Table~\ref{tablobs} --- as well as to
identify the most promising systems to monitor.  This paper supplies
such information.  For a sample of 14 lenses we provide predicted time
delays with uncertainties, a rating of prospects as `excellent',
`good', or `unpromising' based on both models and the observational
situation, and finally an estimate of the precision on $H_0$
obtainable from these lenses.  A companion paper by Eigenbrod et al.\
(2005; COSMOGRAIL I) is devoted to the determination of the optimal
strategy to use in order to measure time-delays (temporal sampling of
the light curves, object visibility and variability, contamination by
microlensing, etc).  Together, these papers help design an
observational campaign.

An ideal time-delay lensing system has the following features: {\bf 1-}
bright optical images, {\bf 2-} large angular image separations
($>$1\arcsec), {\bf 3-} light path unperturbed by nearby structure, {\bf
4-} known or easy-to-measure lens redshift $\zl$.  Our sample of 14 has
been selected using these criteria as a guideline, though not a strict
requirement. We have considered only objects for which the time-delay
can be measured in the optical.

The main results of this paper are in Sect.~\ref{indivsec} and
\ref{predacc}, which present ensembles of models for the 14 individual
systems and then estimate the precision to which $H_0$ could be
recovered from them. But before going into details of the models, it
is useful to preview the results and compare them with measured
systems.  We do this in Sect.~\ref{compsec}.

\section{Comparing observed and predicted delays} \label{compsec}

It is possible to make a preliminary prediction of time delays from
image positions before any modelling, by recalling the scales
involved.

In lensing theory, the geometric part of the time delay is of the
order of the image-separation squared times $DH_0^{-1}$, where $D$ is
the usual dimensionless distance factor depending on
cosmology.\footnote{We refer all time-delay predictions in this paper
to the concordance cosmology ($\Omega_{\rm m}=0.3,
\Omega_\Lambda=0.7$) and $H_0^{-1}=14\,\rm Gyr$ (or $H_0=70$ in local
units).}  The total time delay will be smaller but of the same
order. \cite{s04} shows that the longest time delay can be expressed
as
\begin{equation}
\Delta t = \varphi D \left[{\textstyle\frac1{16}}
                           (\theta_1+\theta_2)^2 \,H_0^{-1}\right]
\label{eq-varphi}
\end{equation}
where where $\theta_1,\theta_2$ are the lens-centric distances (in
radians) of the first and last images to arrive\footnote{In this
section, in order to summarize the time-delays of many lenses, we will
make the brutal simplification of neglecting the second and third
images in quadruples.} and $\varphi$ is a dimensionless factor that
ranges within about 0--2 for quadruples and 2--6 for doubles.  The
expression in square brackets in Eq.~(\ref{eq-varphi}) has the elegant
interpretation of the fraction of the sky covered by the lens, times
$H_0^{-1}$.

We now define an `astrometric time delay' $\Tastrom$ by taking
Eq.~(\ref{eq-varphi}) and setting $\varphi$ to a fiducial value of 1.5
for all quadruples and 4 for all doubles.  This is a useful
preliminary predictor of time delays, as we will see below.

\begin{table}
\caption[]{The 12 time-delays measured so far, with
1$\sigma$ error bars. Lens redshifts in parenthesis 
are either photometric or based on absorption lines in the 
quasar images.}
\label{tablobs}
$$
\begin{array}{p{2cm}p{1cm}cccl}
\hline\noalign{\smallskip}
Object     & Type & \zl      & \Tastrom & & \Tobs \\
\noalign{\smallskip}\hline\noalign{\smallskip}
B0218+357  & AD   &  0.68    &  10 &  & \010\pm1^{\mathrm a,b} \\
J0951+263  & ID   &  (0.24)  &  11 &  & \016\pm2^{\mathrm c} \\
B1115+080  & IQ   &  0.31    &  24 &  & \025\pm4^{\mathrm d,e} \\
B1600+434  & AD   &  0.41    &  35 &  & \051\pm4^{\mathrm f} \\
B0435--122 & CQ   &  0.46    &  41 &  & \014\pm1^{\mathrm g} \\
B1830--211 & AD   &  (0.89)  &  42 &  & \026_{-4}^{+5}{}^{\mathrm h} \\
B2149--274 & AD   &  0.50    &  59 &  & 103\pm12^{\mathrm i} \\
B1608+656  & IQ   &  0.63    &  60 &  & \077\pm3^{\mathrm j} \\
B1520+530  & ID   &  0.72    &  92 &  & 130\pm3^{\mathrm k} \\
J0911+055  & SQ   &  0.77    & 119 &  & 146\pm8^{\mathrm l} \\
B1104--181 & AD   &  0.73    & 345 &  & 161\pm7^{\mathrm m} \\
B0957+561  & ID   &  0.36    & 536 &  & 423\pm1^{\mathrm n} \\
\noalign{\smallskip}\hline
\end{array}
$$
{\rightskip 0pt plus 1cm
$^{\mathrm a}$\cite{cohen00} \hspace*{3mm}  \quad
$^{\mathrm b}$\cite{biggs99} \quad 
$^{\mathrm c}$\cite{jakobsson04} \quad
$^{\mathrm d}$\cite{schechter97} \quad
$^{\mathrm e}$\cite{barkana97}  \hspace*{9mm} \quad
$^{\mathrm f}$\cite{burud00}  \quad
$^{\mathrm g}$\cite{kochanek05}  \quad
$^{\mathrm h}$\cite{lovell98} \quad
$^{\mathrm i}$\cite{burud02a}  \quad
$^{\mathrm j}$\cite{fassnacht02} \quad
$^{\mathrm k}$\cite{burud02b}  \quad
$^{\mathrm l}$\cite{hjorth02} \quad
$^{\mathrm m}$\cite{ofek03}  \quad
$^{\mathrm n}$\cite{oscoz01} \par}
\end{table}

\begin{figure}
\centering \includegraphics[width=.4\textwidth]{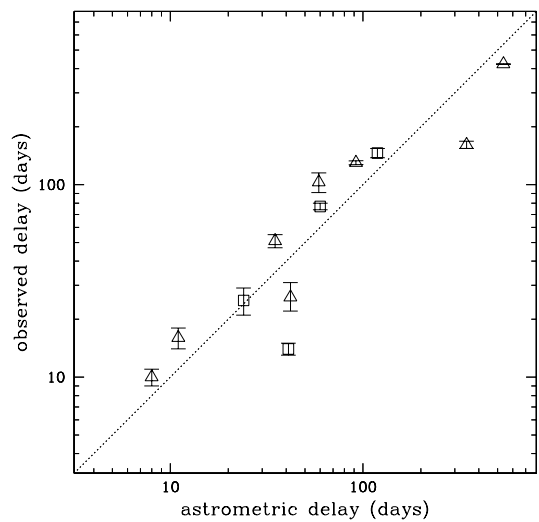}
\caption{Plot of $\Tobs$ against $\Tastrom$ for the known time-delay
  systems, showing that  known  and prospective systems can  be easily
  compared. Squares denote quadruple systems, triangles are for doubles.}
\label{plotobs}
\end{figure}

\begin{table}
\caption[]{Predicted time-delays (and 1$\sigma$ error bars) sorted 
by increasing astrometric  delay, 
for objects with no measured time-delay. Lens redshifts in parenthesis 
are either photometric or based on absorption lines in the 
quasar images.}
\label{tabcand}
$$
\begin{array}{p{2cm}p{1cm}cccl}
\hline\noalign{\smallskip}
Object          & Type & \zl   &  \Tastrom & & \Tpred \\
\noalign{\smallskip}\hline\noalign{\smallskip}
B1422+231         & LQ &  0.34  &   8   &  & 18_{-5}^{+5}  \\
J2026--453        & IQ & (0.5)  &  14   &  & 15_{-6}^{+2}  \\
J1155+634         & AD &  0.18  &  19   &  & 35_{-10}^{+8} \\
J0924+021         & IQ &  0.39  &  19   &  & 12_{-4}^{+6} \\
J1650+425         & ID & (0.5)  &  46   &  & 54_{-13}^{+8} \\
J1335+011         & AD &  0.44  &  47   &  & 49_{-16}^{+13} \\
J1355--225        & AD & (0.70) &  68   &  & 89_{-39}^{+28} \\
J1131--123        & LQ &  0.30  &  69   &  & 137_{-39}^{61} \\
J2033--472        & IQ &  0.66  &  70   &  & 72_{-20}^{+33} \\
B1030+074         & AD &  0.60  &  75   &  & 153_{-57}^{+29}\\
B0909+532         & ID & (0.83) &  90   &  & 72_{-17}^{+10} \\
B1009--025        & AD &  0.87  &  98   &  & 161_{-59}^{+34}\\
B0818+122         & ID &  0.39  & 111   &  & 110_{-26}^{+16}\\
J0903+502         & ID &  0.39  & 122   &  & 110_{-23}^{+13}\\
\noalign{\smallskip}\hline
\end{array}
$$
\end{table}

\begin{figure}
\centering \includegraphics[width=.4\textwidth]{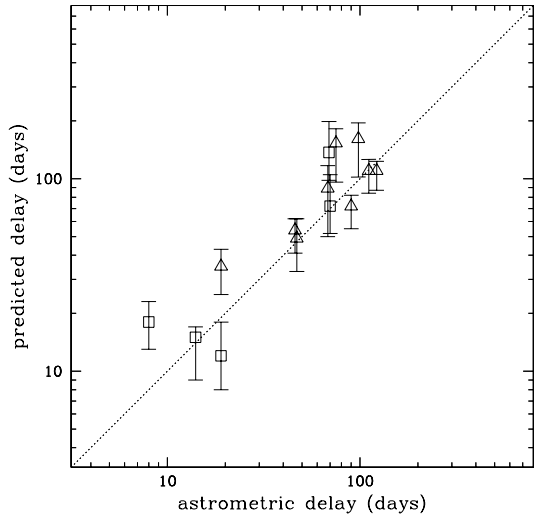}
\caption{Plot of $\Tpred$ against $\Tastrom$ for the prospective
time-delay systems. Error bars are 68\% confidence. Squares denote
quadruple systems, triangles are for doubles.}
\label{plocand}
\end{figure}

\begin{figure}
\centering \includegraphics[width=.4\textwidth]{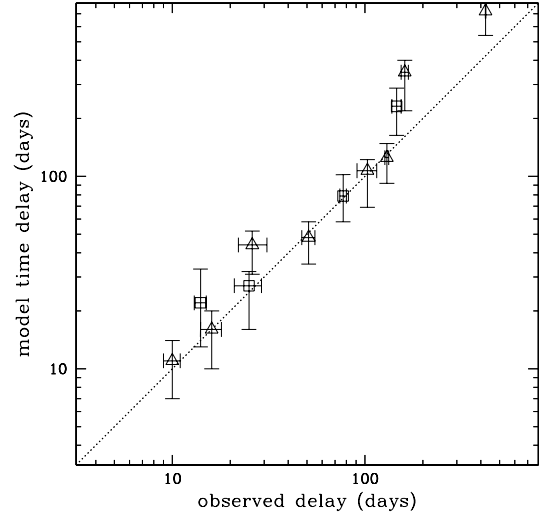}
\caption{Plot of $\Tpred$ against $\Tobs$ for the current
time-delay systems. Again, squares are for quadruple systems,
triangles for doubles.}
\label{plotest}
\end{figure}

Table~\ref{tablobs}  gives the  astrometric and  actual  observed time
delays  for the 11   known time-delay systems  (disregarding here  the
middle two images in quadruples, i.e., images 2 and 3 in the figures).
The `type'  refers  to the morphological classification  introduced in
\cite{sw03}: $\hbox{AD}  = \hbox{axial}\ \hbox{double,}$  $\hbox{ID} =
\hbox{inclined}\   \hbox{double,}$   $\hbox{CQ}     =     \hbox{core}\ 
\hbox{quad,}$ $\hbox{LQ}  = \hbox{long-axis}\ \hbox{quad,}$ $\hbox{SQ}
=  \hbox{short-axis}\  \hbox{quad,}$  $\hbox{IQ} =    \hbox{inclined}\ 
\hbox{quad.}$

Fig.~\ref{plotobs} plots the data summarized in Table~\ref{tablobs}.
It is striking that while $\Tobs$ ranges over a factor of 40, it
tracks $\Tastrom$ to a factor of~2.

Fig.~\ref{plocand}  and Table~\ref{tabcand}   summarize our time-delay
predictions.  To  make these predictions  we used the {\em PixeLens\/}
code (\cite{sw04})  to generate  an ensemble of  200  models  for each
lens,  leading to an ensemble  of model time-delays, which we interpret
as the probability distribution for the predicted time-delays.

How reliable are the time-delay predictions?  Pixellated models
generically involve a choice of prior (also called secondary
constraints); if the prior is too different from what lenses are
really like then the results will be incorrect.  Our prior is
basically the {\em PixeLens\/} default; in detail, we assumed the
following:

\begin{enumerate}

\item In most cases we required the mass profile to be
  inversion-symmetric about the lens centre.  But if the lensing galaxy
  appeared very asymmetric, or the image morphology was unusual, we let
  the mass profile be asymmetric.

\item If there was evidence of external shear from the lens environment
  and/or the image morphology, we allowed the code to fit for constant
  external shear.  That is to say, we allowed a contribution of the form
  $\gamma_1(\theta_x^2-\theta_y^2)+2\gamma_2\theta_x\theta_y$ to the
  arrival time, with adjustable constants $\gamma_1,\gamma_2$.

\item The density gradient must point within $45^\circ$ of the lens
  center (thus ensuring that the lens is centrally concentrated).

\item The radial mass profile must be steeper than $\theta^{-0.5}$.
That implies a 3D profile steeper than $r^{-1.5}$, which is consistent
with available estimates from stellar or gas dynamics; for example,
\cite{binney91} report an $r^{-1.75}$ profile near the Galactic centre.

\item The density on any pixel must be $\leq$ twice the average of its
  neighbours, except for the central pixel, which can be arbitrarily
  dense.

\end{enumerate}

As a test we `postdicted'  the time-delays in  the known systems.  The
results are summarized in Fig.~\ref{plotest}.  We  find that our prior
tends to overestimate the time-delays for the systems with the largest
angular separations,  perhaps because these  lenses have a significant
cluster contribution  and the profiles are much  shallower than in our
prior. One  of   the discrepant  lenses is PKS~1830-211,   which has a
double  lens galaxy. The  two others are  B0957$+$561 and J0911$+$055,
which both have  significant contribution  by  a group  or cluster  of
galaxies along  the line of sight.   But predicted time delays of less
than 200 days  appear reliable.  The candidate lenses  are all  in the
reliable regime.

\section{Individual systems} \label{indivsec}

We now proceed to discuss individual lenses, grouped by similar
morphology.

For each lens, we show three kinds of plot.  First, there is a mass
map of the ensemble-average model.  The contours in the mass maps are
in logarithmic steps, with each step corresponding to a factor of
$10^{0.4}$ (like a magnitude scale).  The critical density contour is
always the third from outermost.  Second, we have plots showing
saddle-point contours.  These plots also show the source position in
the ensemble-average model.  The detailed placement of the
saddle-point contours and the inferred source varies across the
ensemble, but the qualitative features are robust.  In particular, the
saddle-point contours make the time-ordering of images obvious.  We
will refer to individual images by their time order: 1,2 for doubles
or 1,2,3,4 for quadruples, meaning that the image labelled~1 varies
first, then 2, etc.  Third, we have histograms for the predicted time
delays between different pairs of images in each lens.

After modelling each lens, we rate its prospects as a time-delay
system as `excellent', `good', or `unpromising', based on how
well-constrained the time-delays are and on the comparative ease of
monitoring and photometry.

We remark that the modelling process really produces a predicted
distribution for $H_0\,\Delta t$.  In the present work we insert a
fiducial value of $H_0$ to obtain a distribution for $\Delta t$, but one
can equally insert a measured value of $\Delta t$ (if available) and
obtain a distribution for $H_0$.  But if two or more time delays become
available for a quadruple, their ratio provides a new constraint on the
lens, and the modelling code must be run again.


\subsection{Axial doubles}

\begin{figure}
\centering
\includegraphics[width=.15\textwidth]{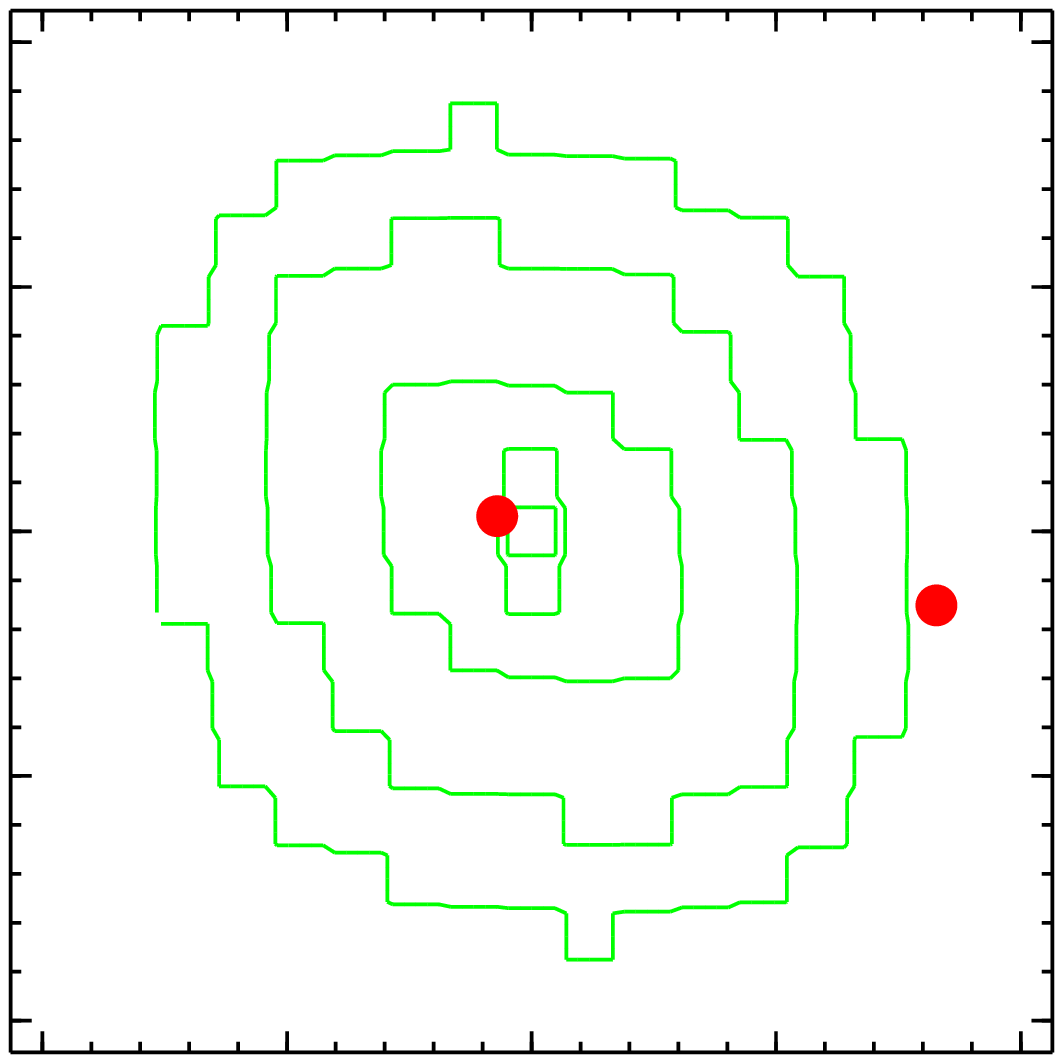}%
\includegraphics[width=.15\textwidth]{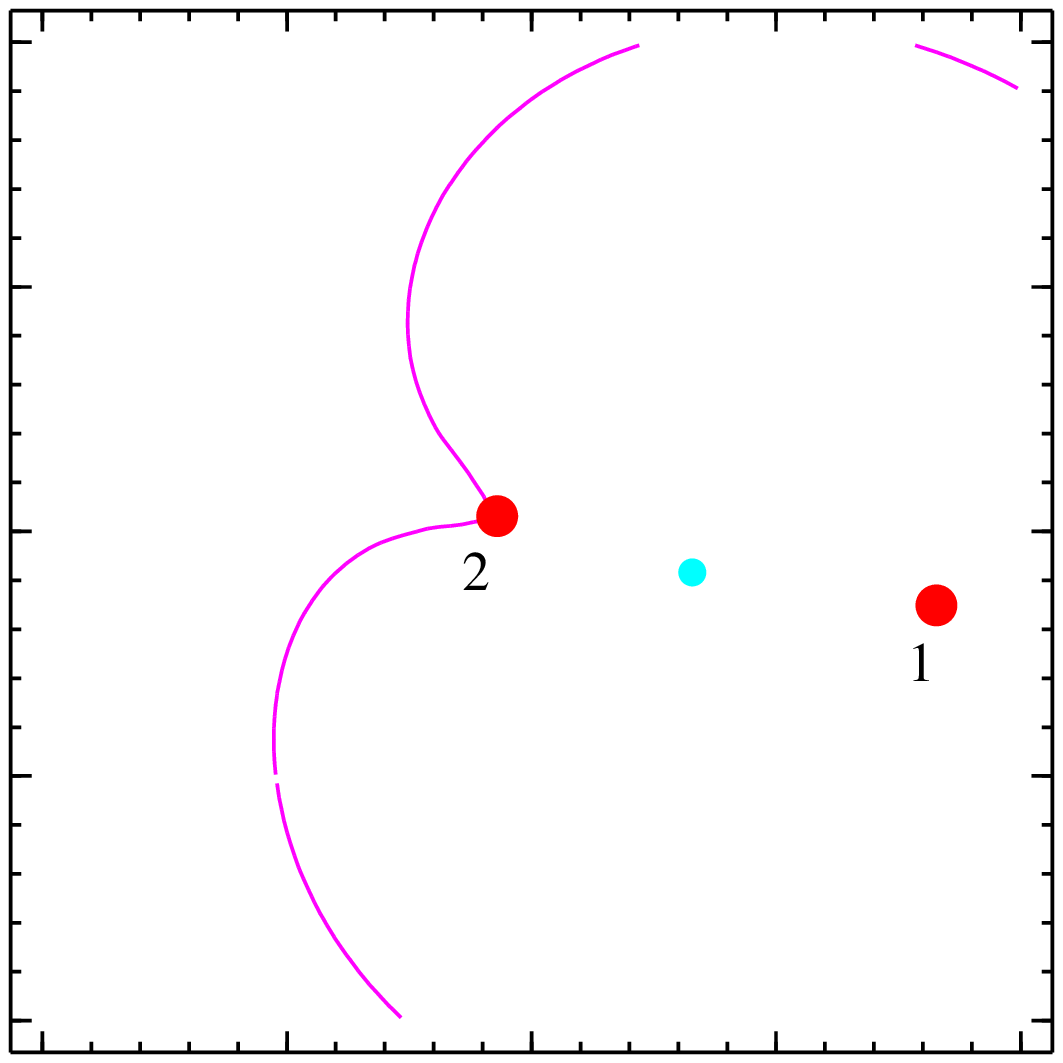}
\figskip
\includegraphics[width=.3\textwidth]{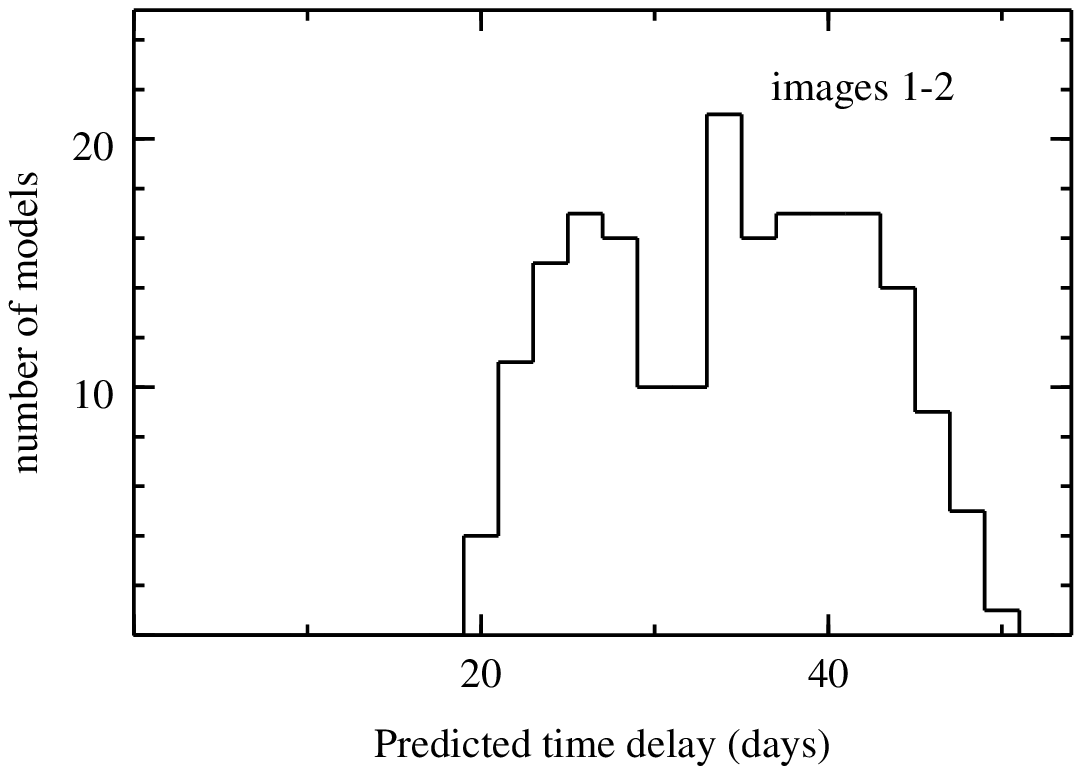}
\caption{Models of J1155+634 (axial double).  See text in
section~\ref{indivsec} for the format. Prospects:
unpromising.}
\label{plotJ1155}
\end{figure}

\noindent {\bf J1155+634}~[Fig.~\ref{plotJ1155}] discovery:
\cite{pindor04}. The separation $\Delta\theta =1.83''$ is relatively
large, but the lens galaxy is only $\sim0.2''$ from the fainter image.
Also, the measurement $\zl=0.1756$ is somewhat insecure because the
inferred galaxy absorption features are amongst the Ly$\alpha$ forest
lines.  As a time-delay prospect, this system appears unpromising.

\begin{figure}
\centering
\includegraphics[width=.15\textwidth]{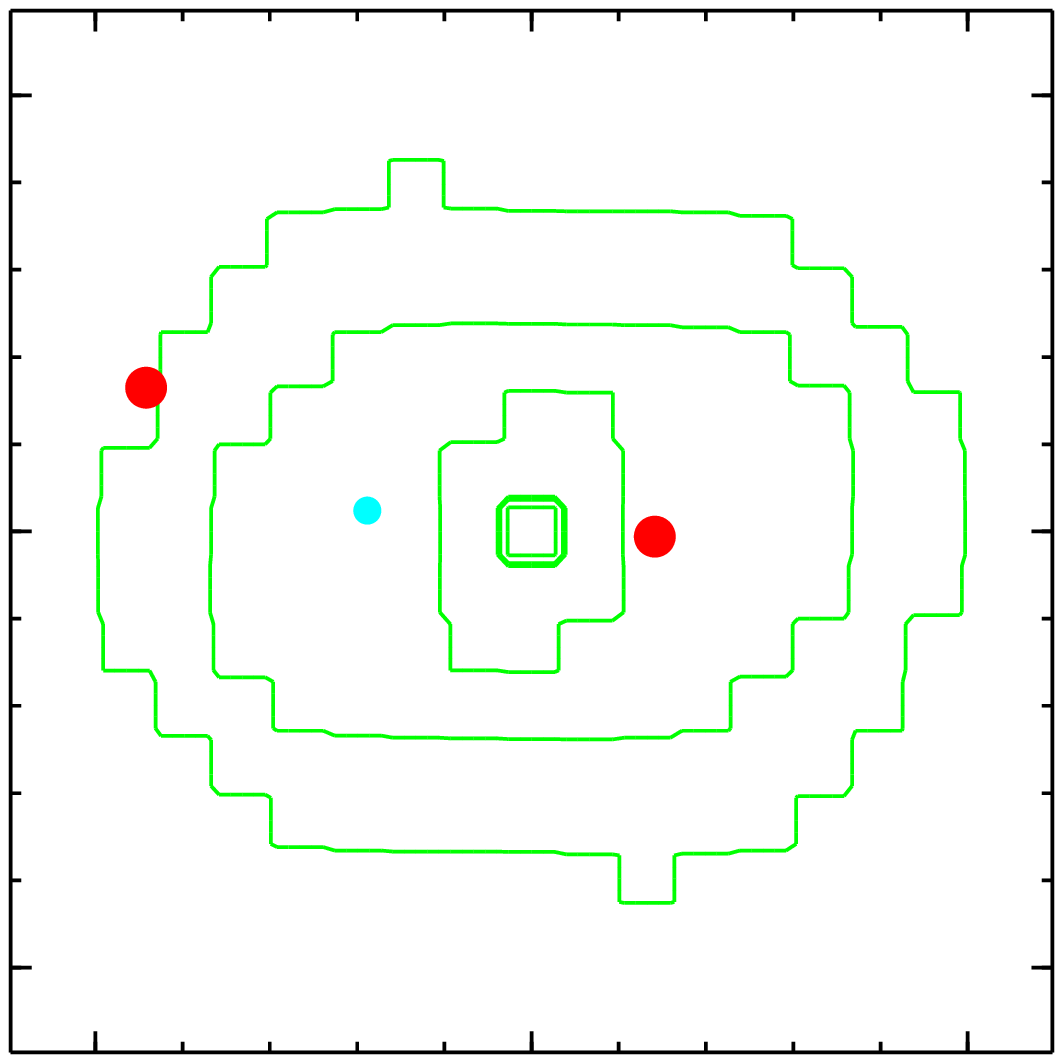}%
\includegraphics[width=.15\textwidth]{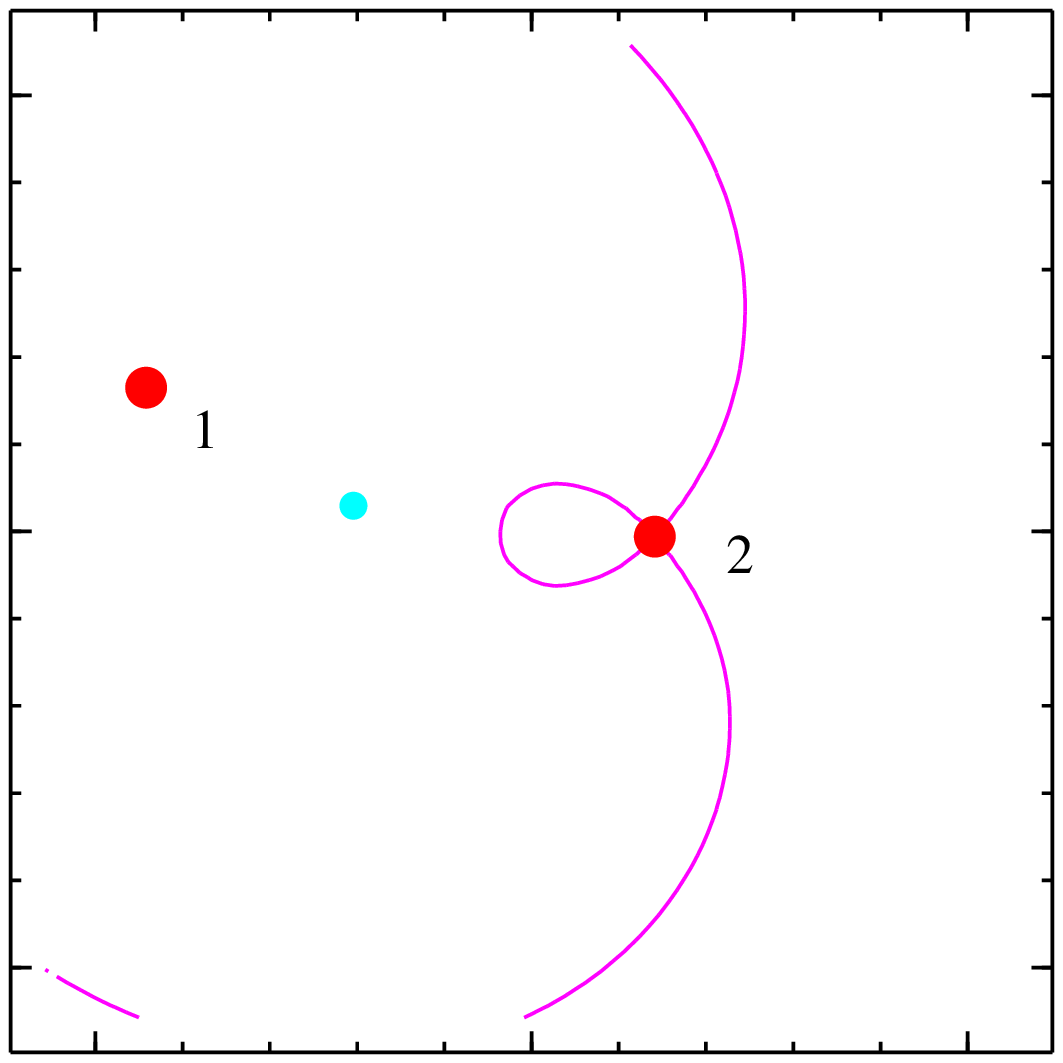}
\figskip
\includegraphics[width=.3\textwidth]{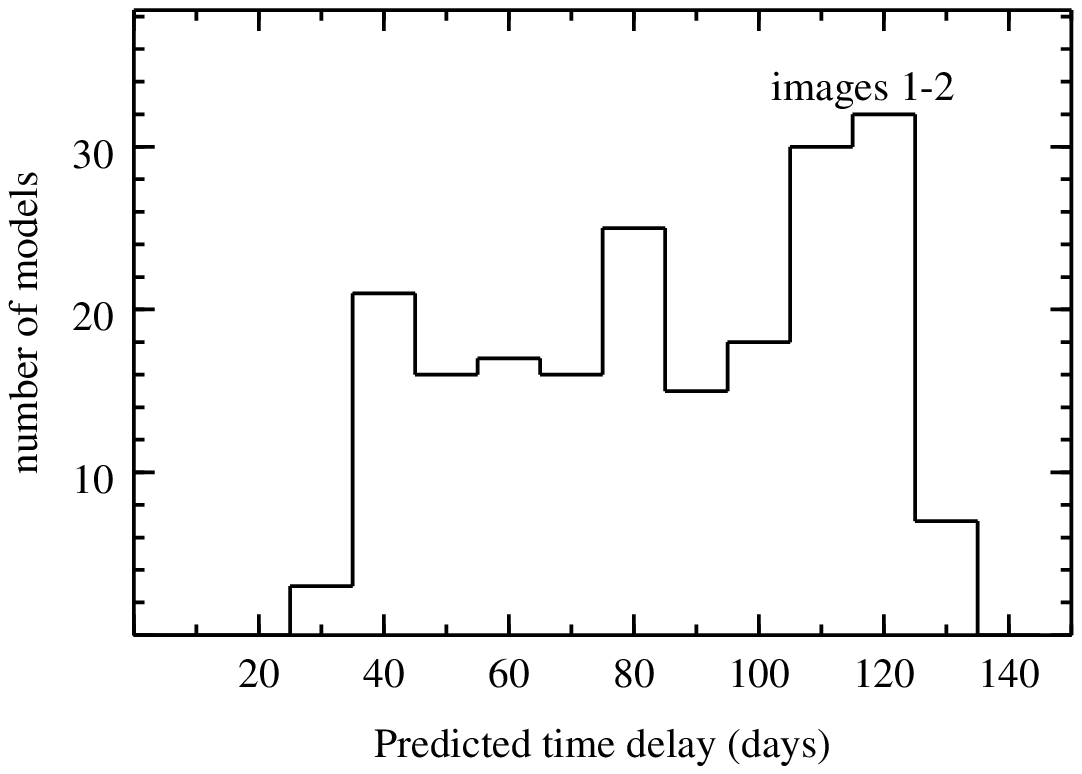}
\caption{Models of J1355--225 (axial double). Prospects: good.}
\label{plotJ1355}
\end{figure}

\noindent {\bf J1355--225} [Fig.~\ref{plotJ1355}] discovery:
\cite{morgan03a}; also known as CTQ~327.  The quasar images are bright
and the angular separation is moderate: $\Delta\theta =1.22''$.
Models include external shear corresponding to further mass to the NW
or SE.  We rate this system as a good time-delay prospect.

\begin{figure}
\centering
\includegraphics[width=.15\textwidth]{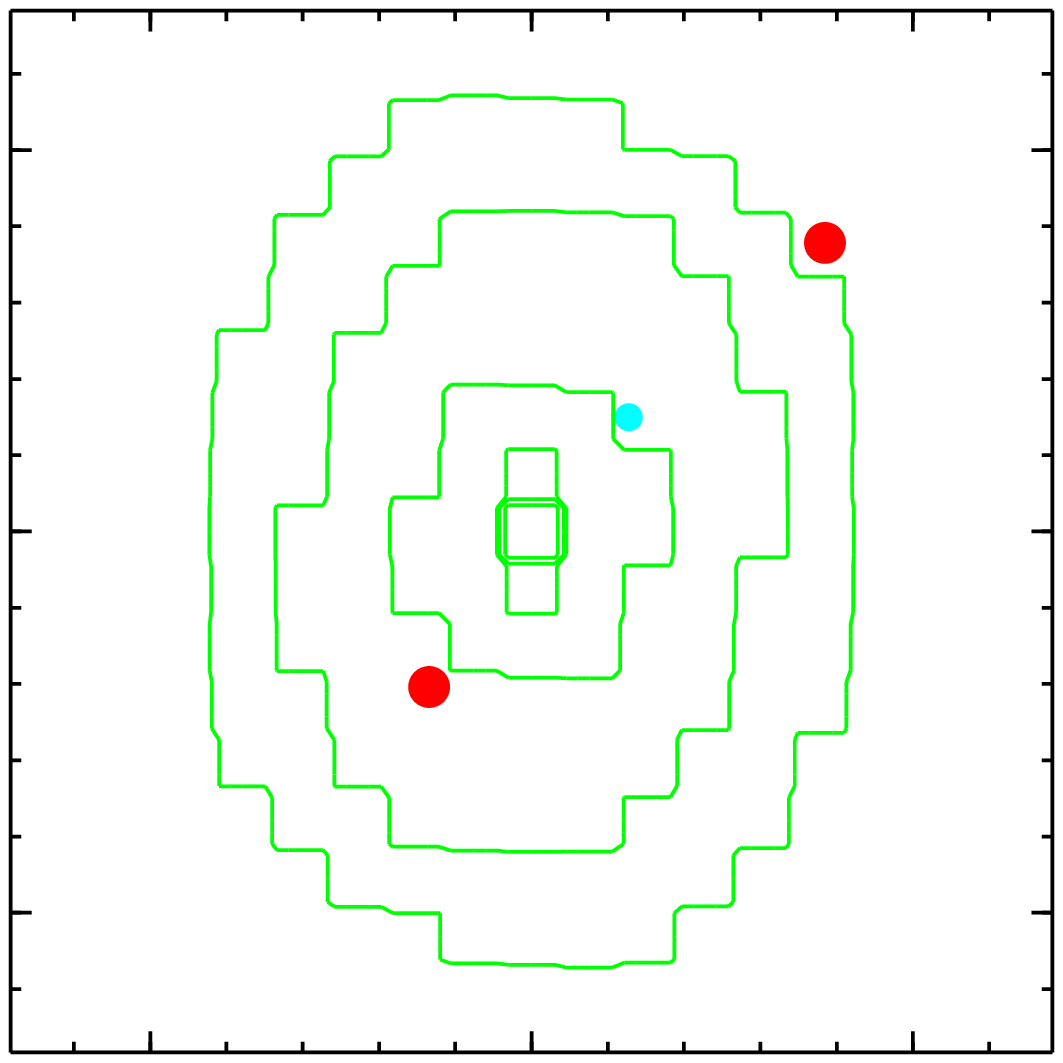}%
\includegraphics[width=.15\textwidth]{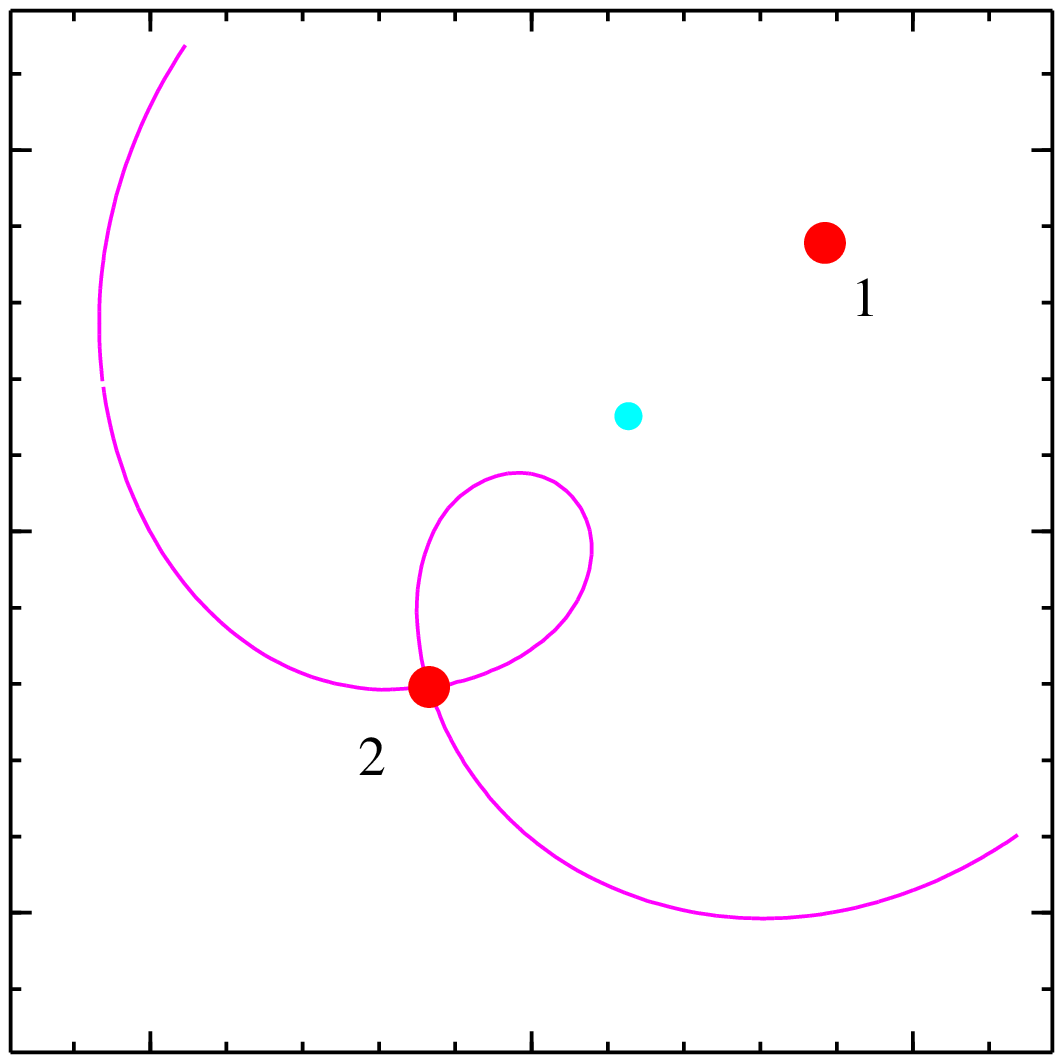}
\figskip
\includegraphics[width=.3\textwidth]{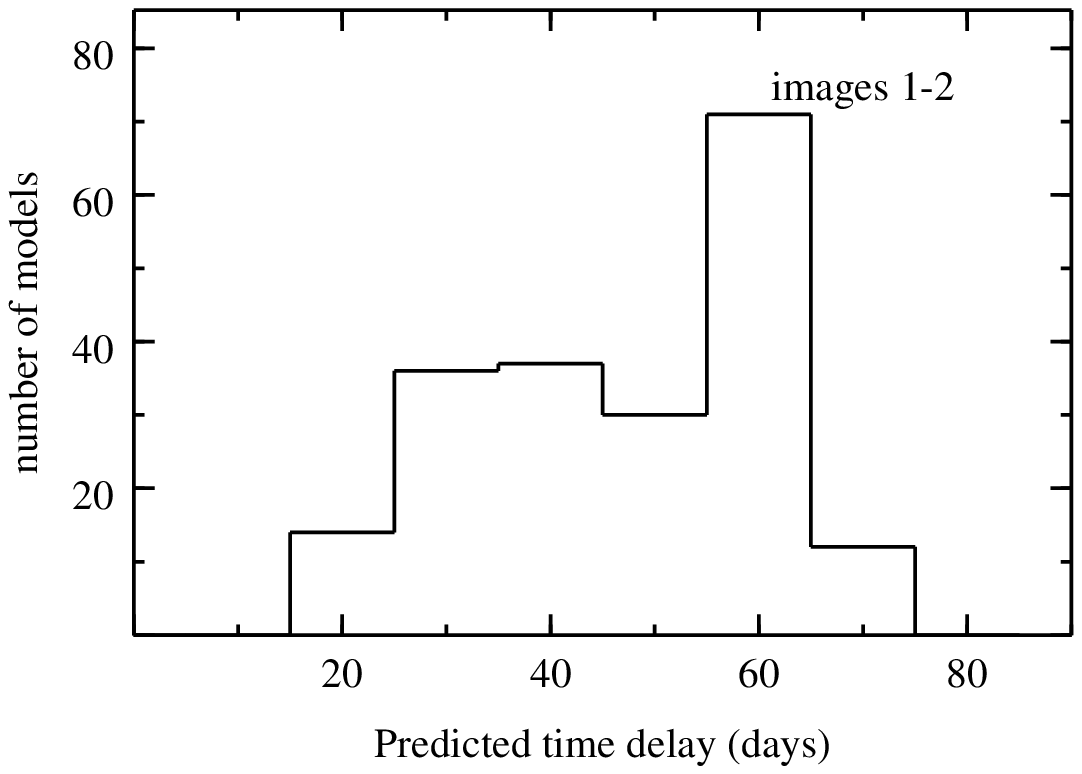}
\caption{Models of J1335+011  (axial double). Prospects: excellent.}
\label{plotJ1335}
\end{figure}

\noindent {\bf J1335+011} [Fig.~\ref{plotJ1335}] discovery:
\cite{oguri04b}.  We rate this system as an excellent time-delay
prospect.

\begin{figure}
\centering
\includegraphics[width=.15\textwidth]{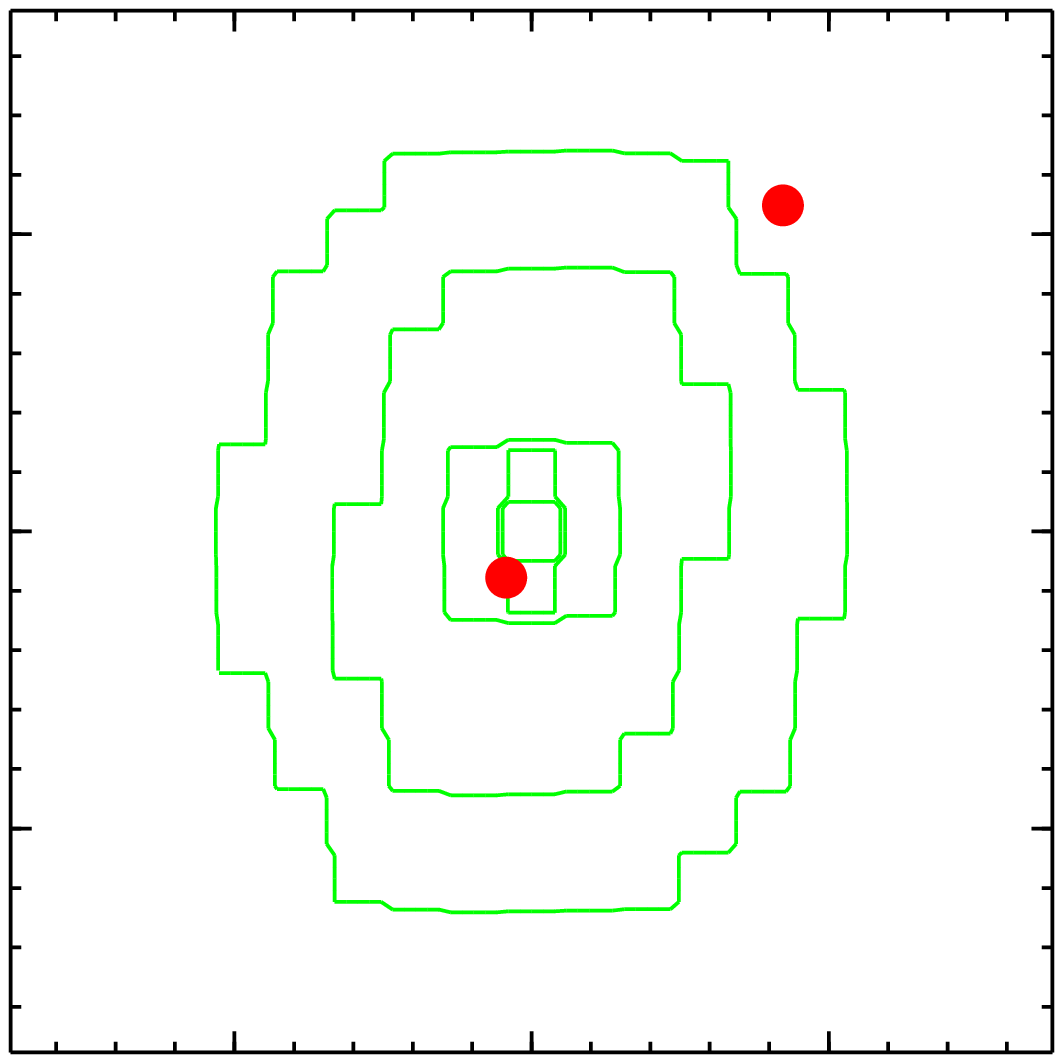}%
\includegraphics[width=.15\textwidth]{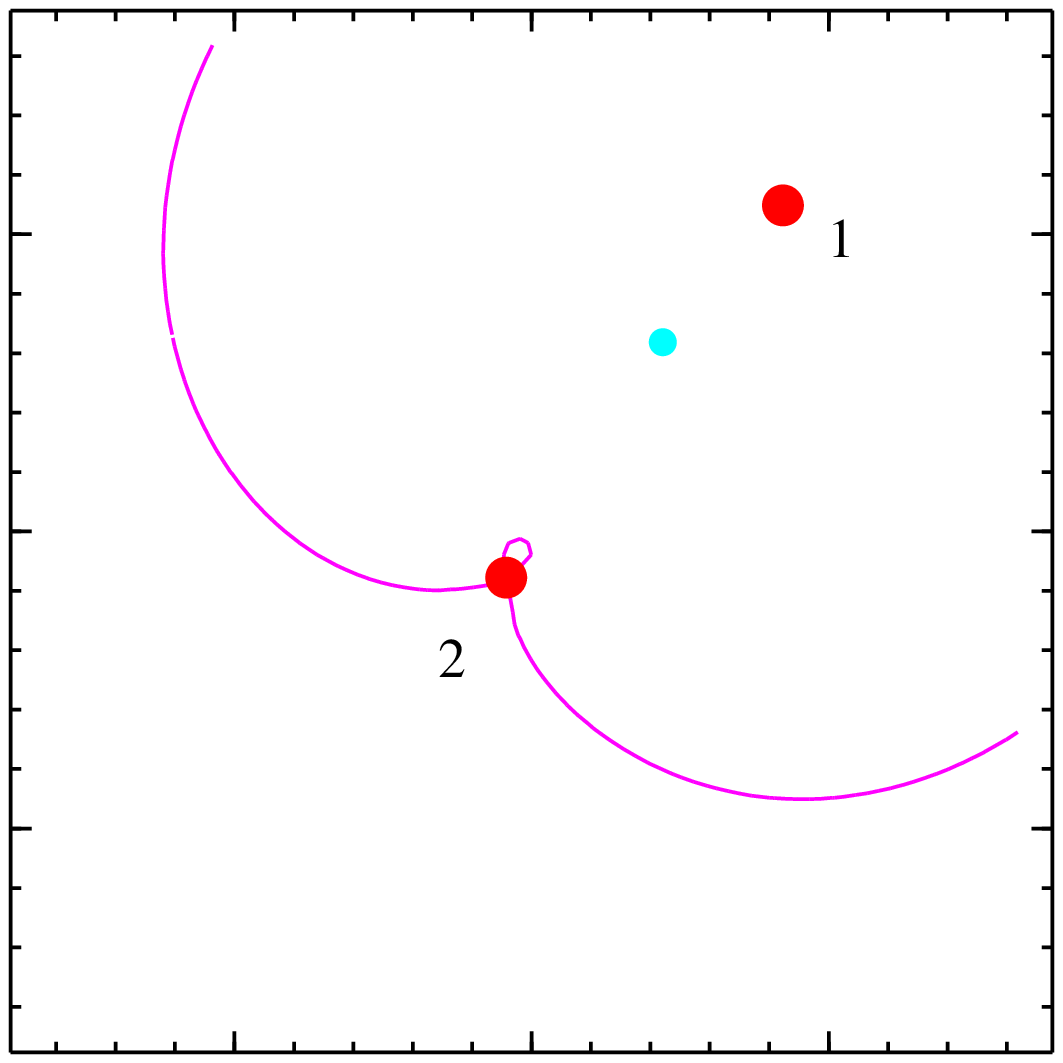}
\figskip
\includegraphics[width=.3\textwidth]{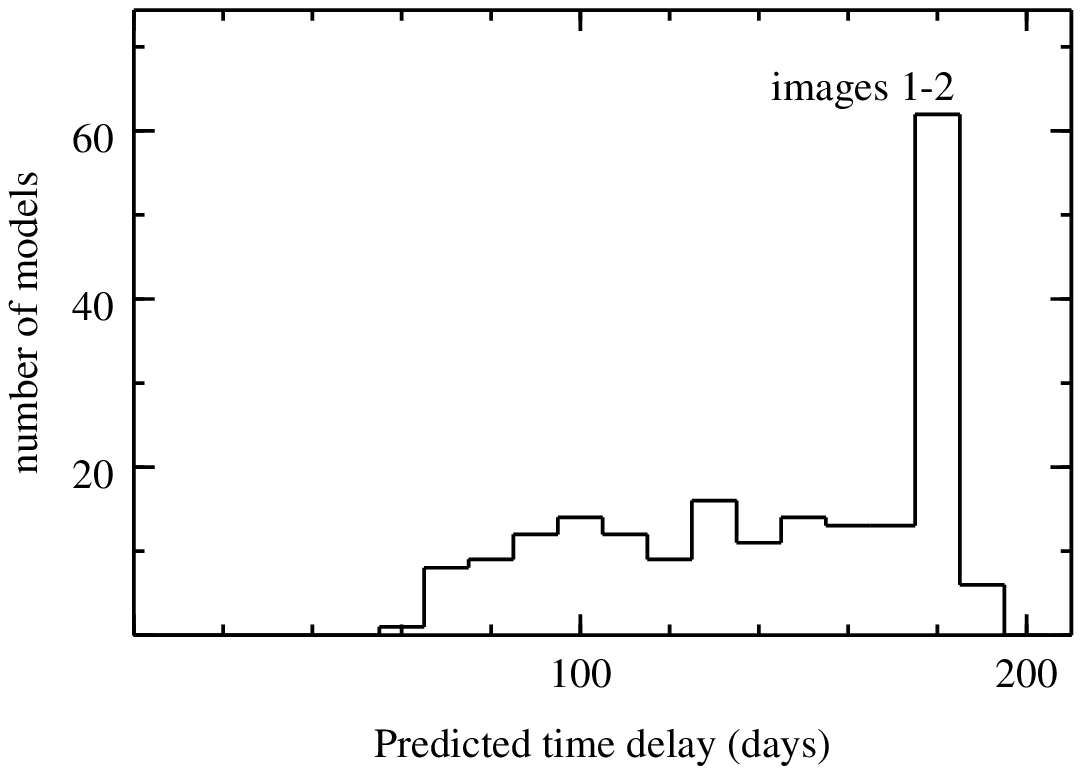}
\caption{Models of B1030+074 (axial double). Prospects:
unpromising.}
\label{plotB1030}
\end{figure}

\noindent {\bf B1030+074} [Fig.~\ref{plotB1030}] discovery:
Xanthopoulos et al.\ (1998).  Like J1155+634 it has a relatively wide
separation but a second image is faint and very close to the galaxy.
There is evidence for variability. The peak in the predicted time
delays near 180\thinspace days is interesting, but it is probably not
wise to over-interpret, given the resolution of the models used in
this paper.  Because of the difficulty of accurate photometry on the
second image, we rate this system as unpromising.

\begin{figure}
\centering
\includegraphics[width=.15\textwidth]{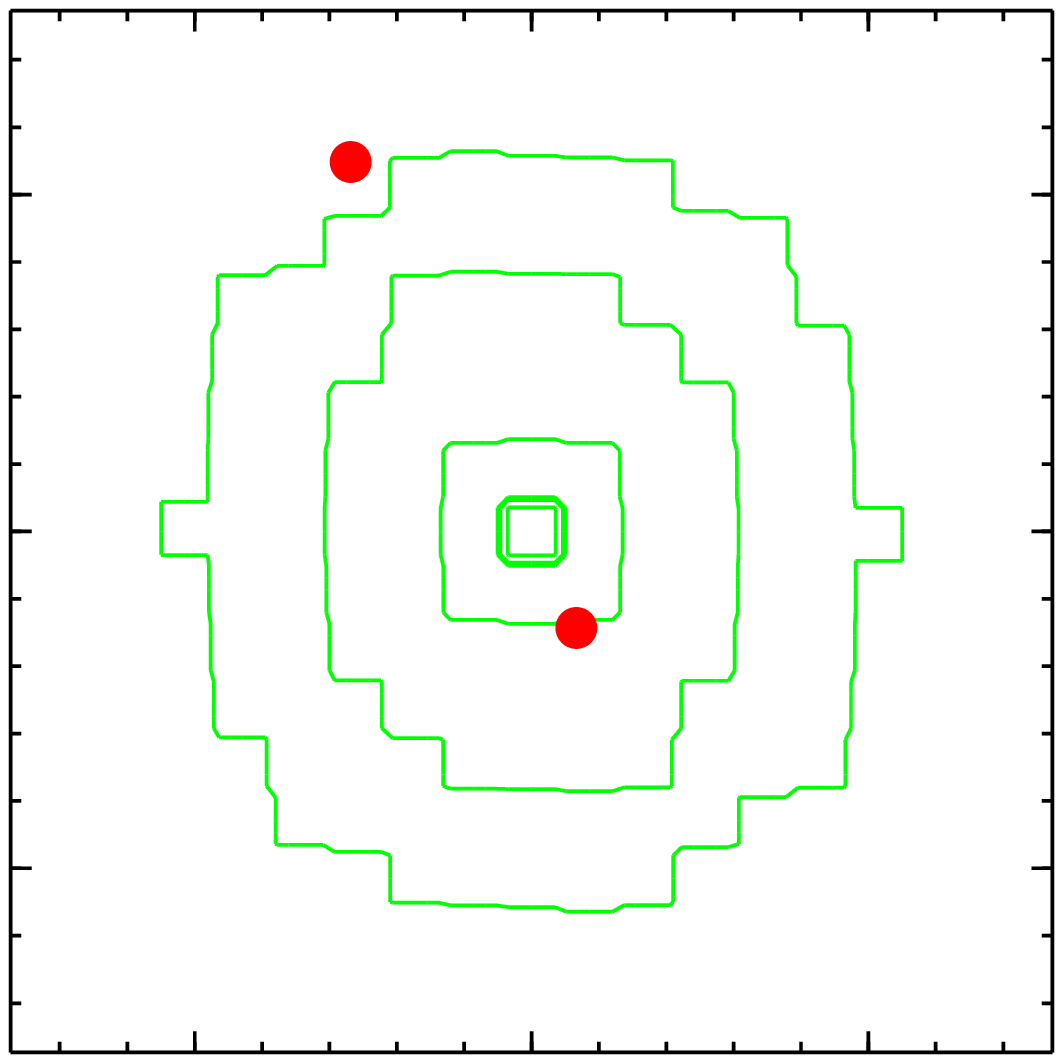}%
\includegraphics[width=.15\textwidth]{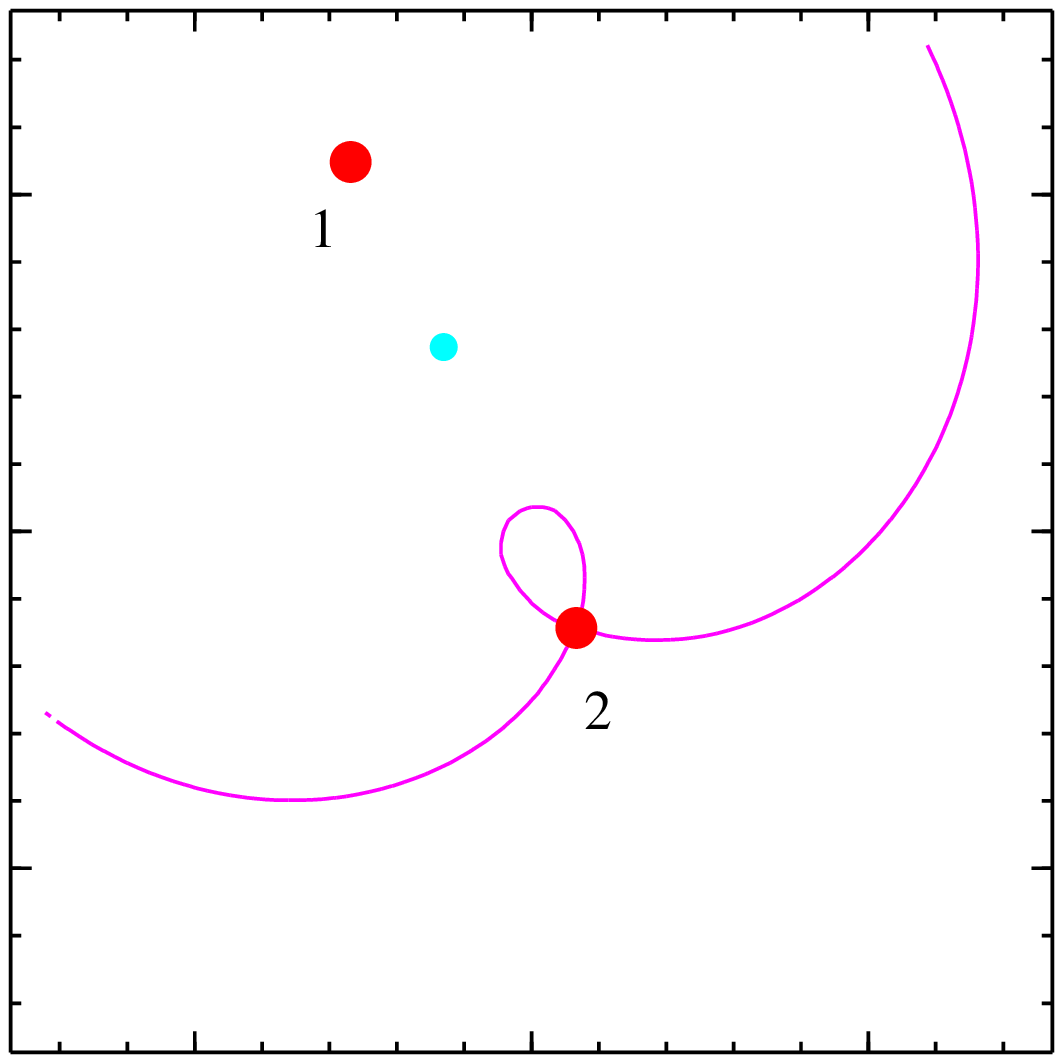}
\figskip
\includegraphics[width=.3\textwidth]{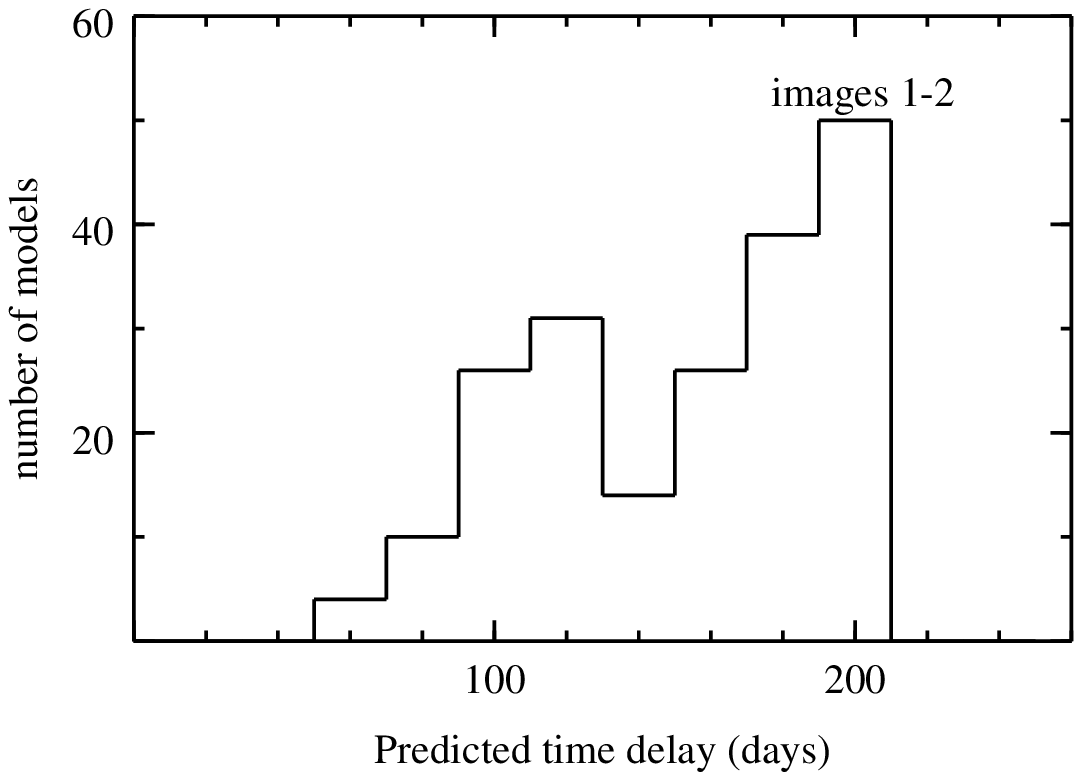}
\caption{Models of B1009--025 (axial double). Prospects:
good.}
\label{plotB1009}
\end{figure}

\noindent {\bf B1009--025} [Fig.~\ref{plotB1009}] discovery:
\cite{surdej93}.  Its clean morphology, evidence of variability and a
nearby foreground QSO usable as a standard PSF all make this an
attractive target.  However, the combination of an approximately
half-year time-delay and a near-equatorial location is awkward (see
Eigenbrod et al.  2005 for more details).  We rate time-delay
prospects as good.


\subsection{Inclined doubles}

\begin{figure}
\centering
\includegraphics[width=.15\textwidth]{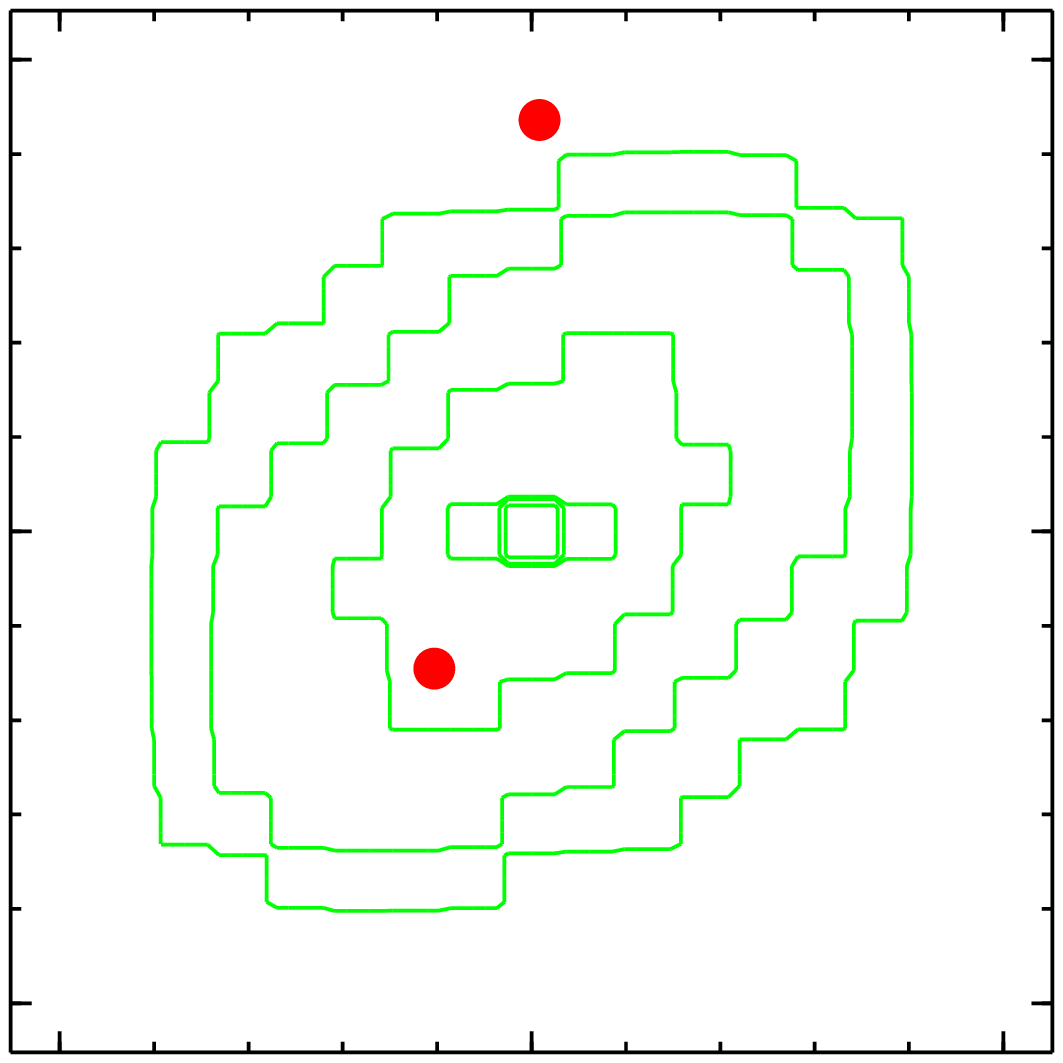}%
\includegraphics[width=.15\textwidth]{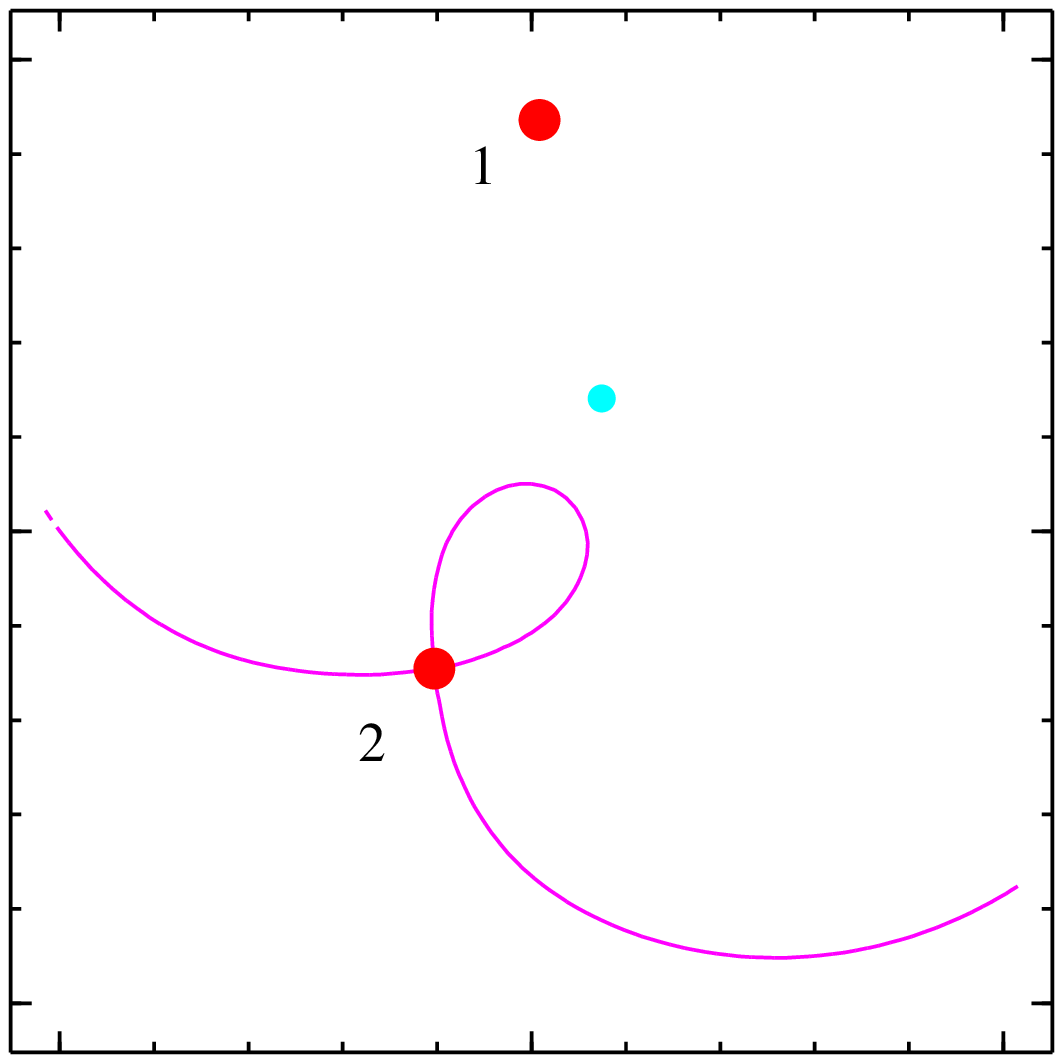}
\figskip
\includegraphics[width=.3\textwidth]{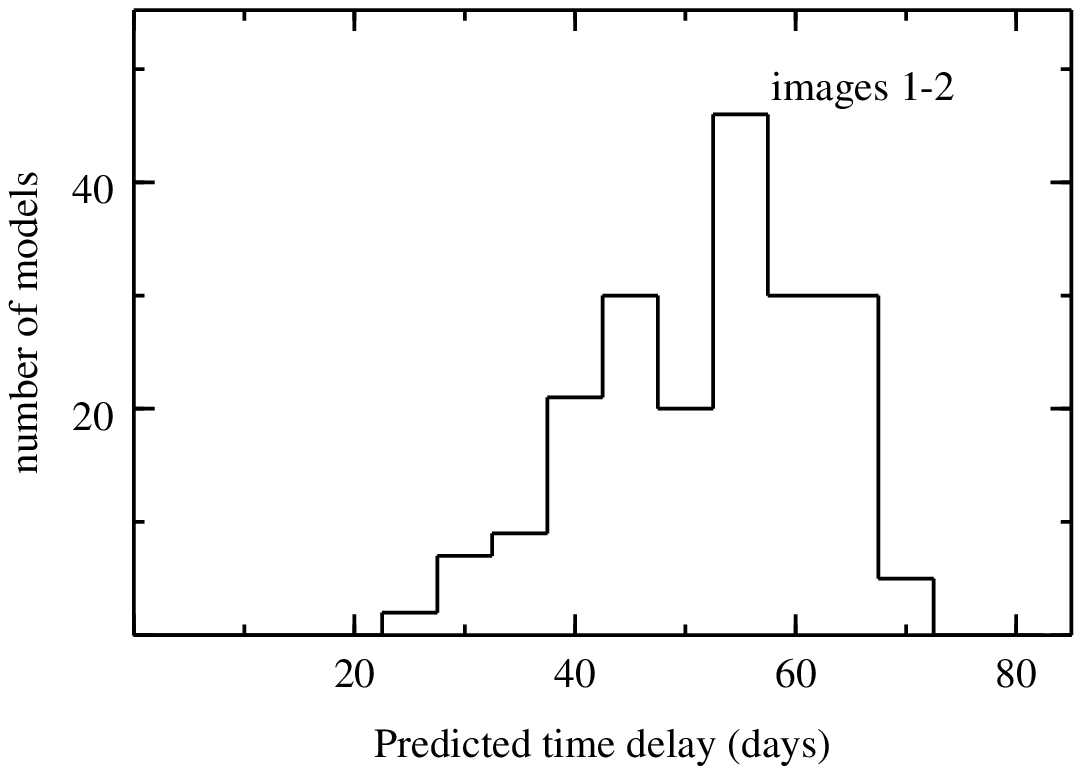}
\caption{Models of J1650+425 (inclined double). Prospects: excellent.}
\label{plotJ1650}
\end{figure}

\noindent {\bf J1650+425} [Fig.~\ref{plotJ1650}] discovery:
\cite{morgan03b}.  It has significant external shear, probably from a
group galaxy to the E.  The high declination of the objects makes it
possible to observe it almost continuously from the northern
hemisphere.  This system is an excellent time-delay prospect.

\begin{figure}
\centering
\includegraphics[width=.15\textwidth]{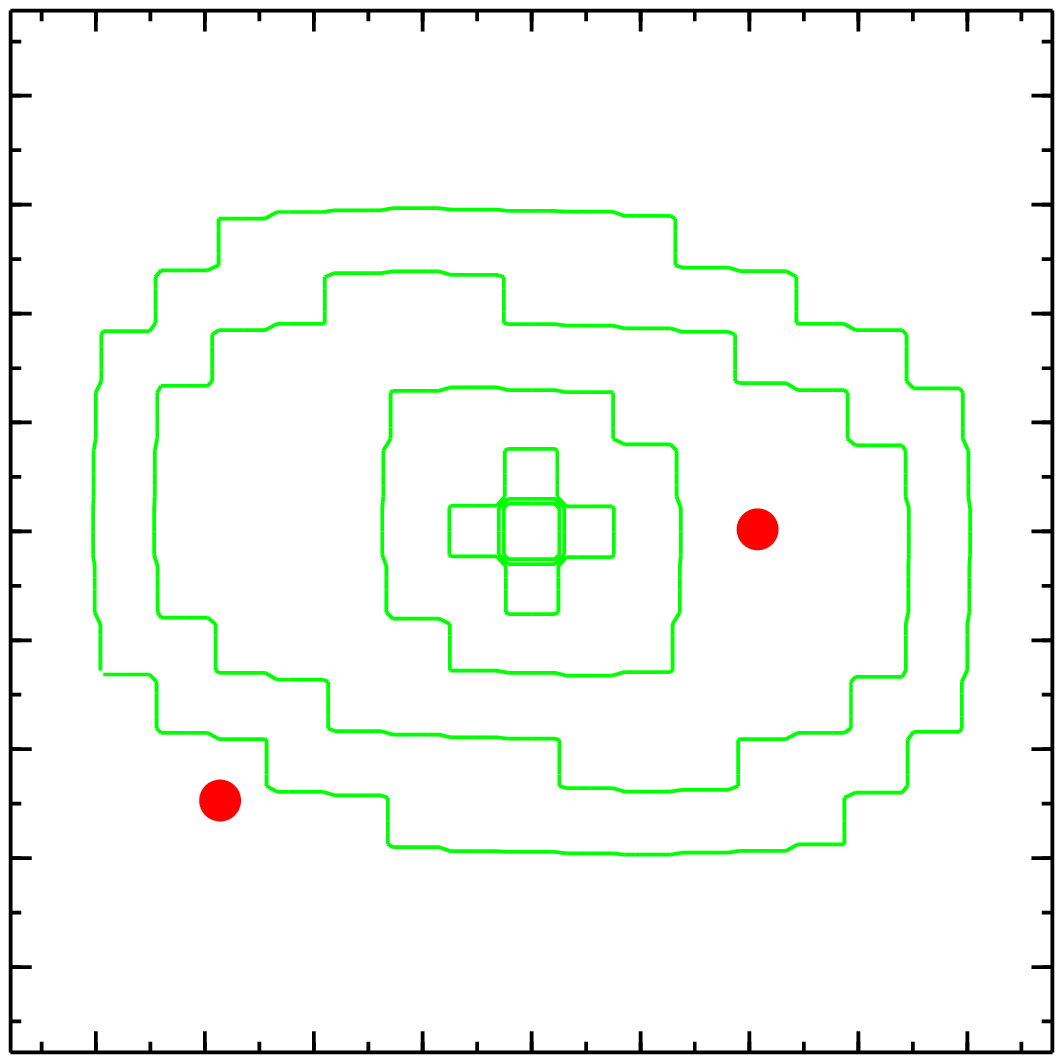}%
\includegraphics[width=.15\textwidth]{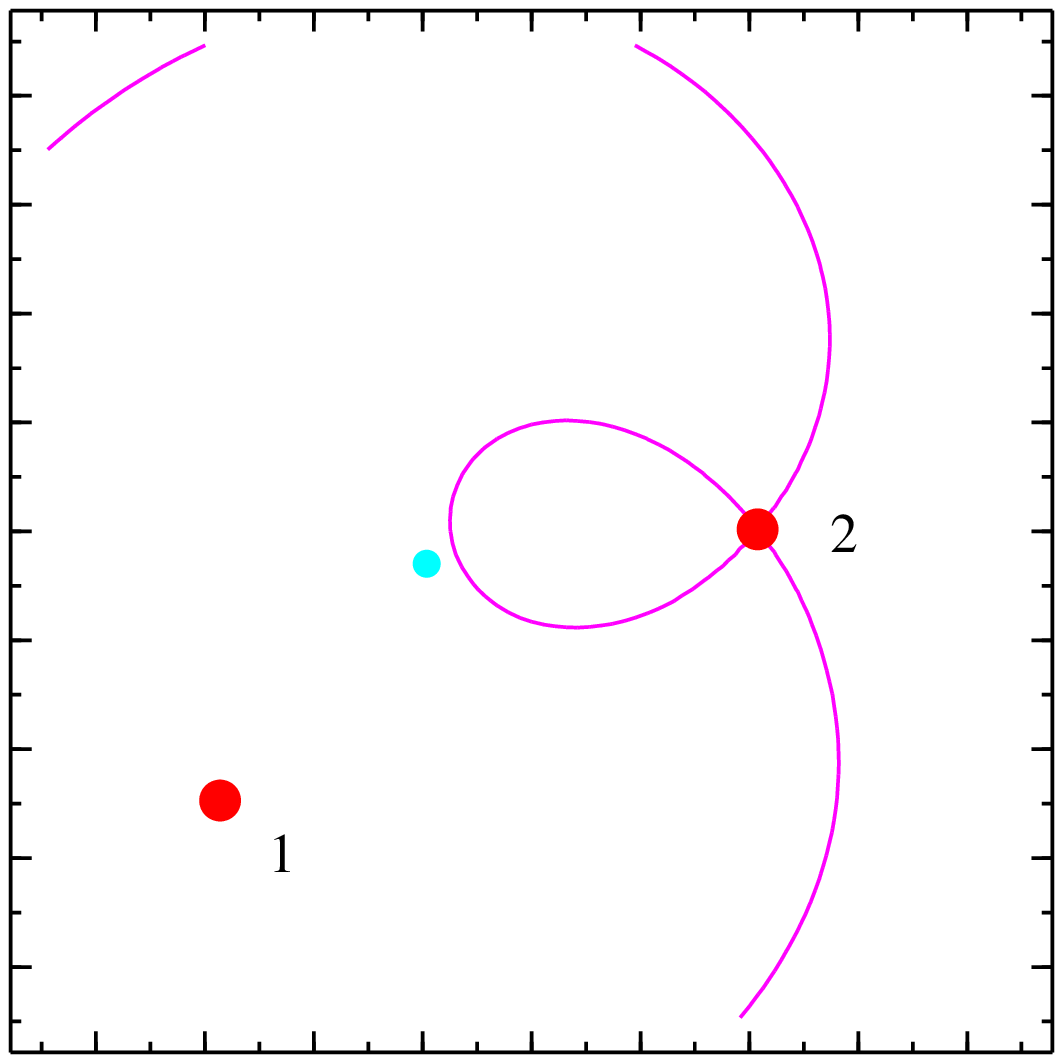}
\figskip
\includegraphics[width=.3\textwidth]{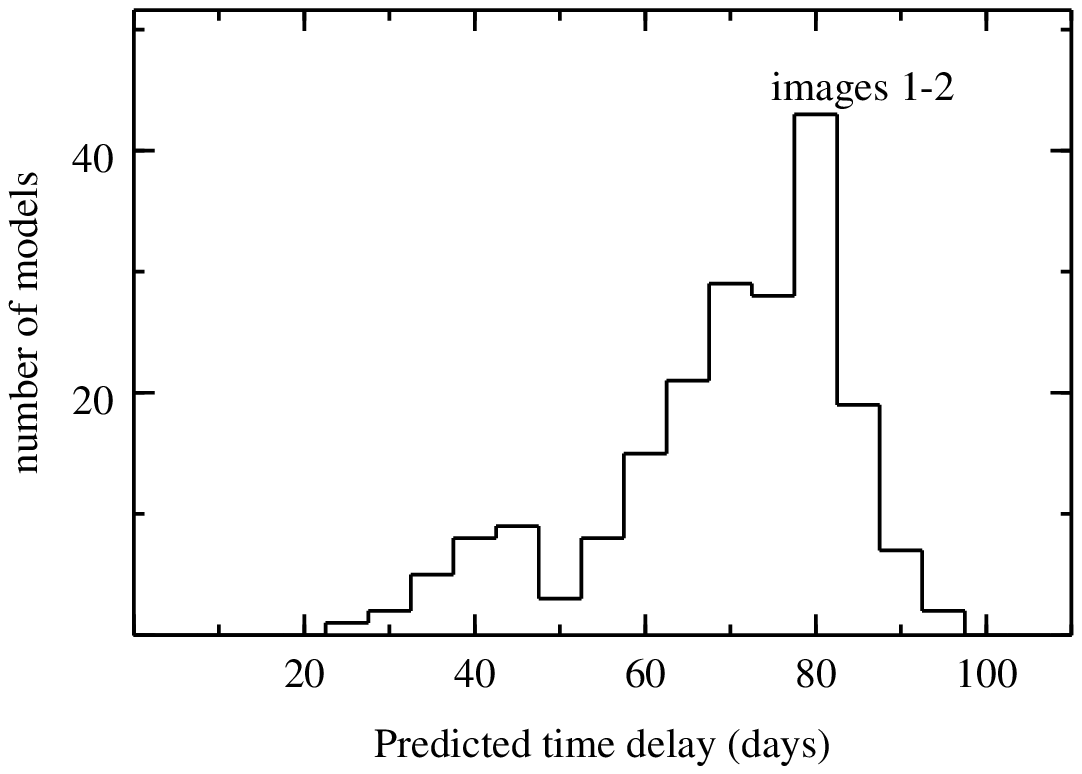}
\caption{Models of B0909+532 (inclined double). Prospects: excellent}
\label{plotB0909}
\end{figure}

\noindent {\bf B0909+532} [Fig.~\ref{plotB0909}] discovery as multiply
imaged: \cite{kochanek97}. The lensing galaxy is very faint, which caused
some early controversy until the issue was settled by \cite{oscoz97} and
\cite{lubin00}.  The morphology and models indicate significant external
shear from mass to the NE or SW, but the galaxies responsible have not
yet been identified. Both quasar images are very bright, and their
separation is $\Delta\theta =1.17''$.  Its $\zl$ is still insecure, but
assuming that problem is solved, this system is an excellent prospect.

\begin{figure}
\centering
\includegraphics[width=.15\textwidth]{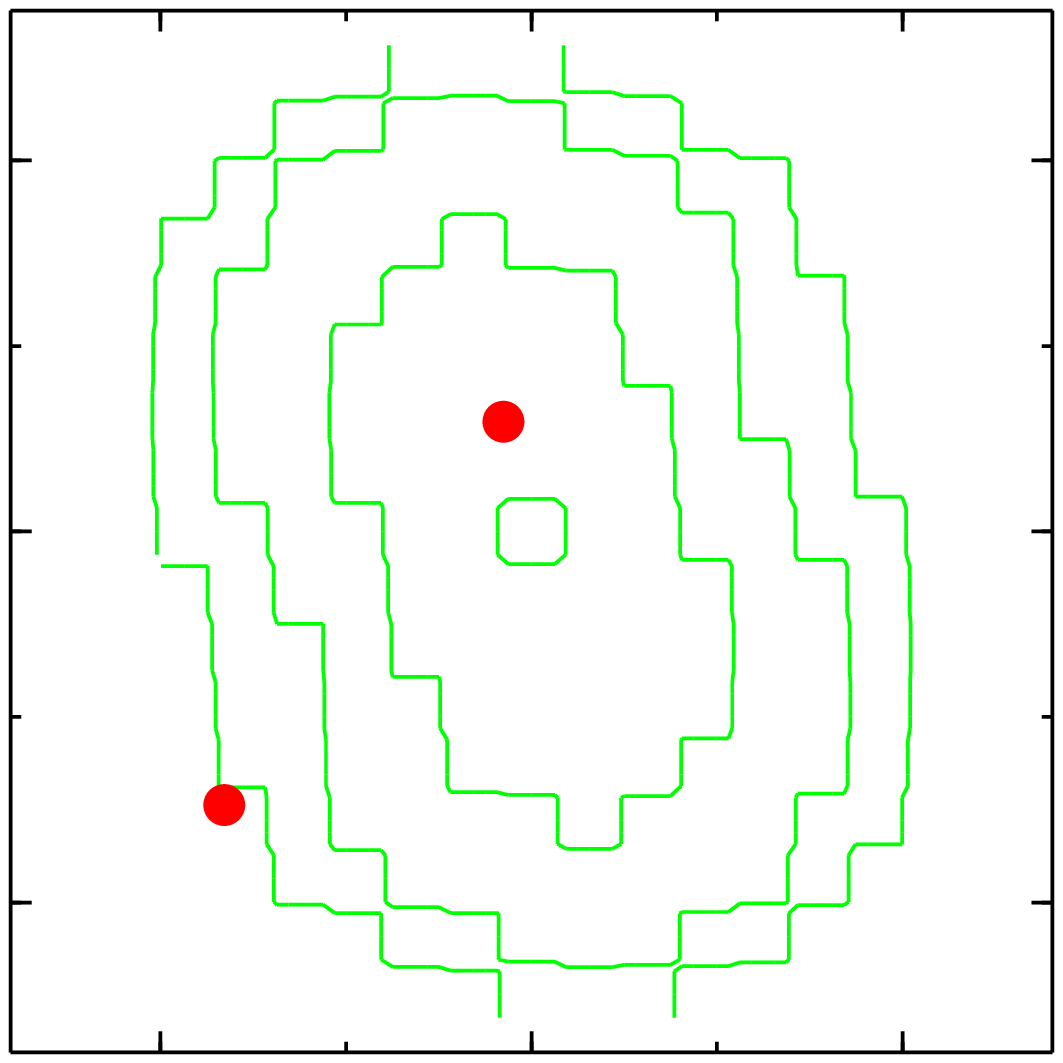}%
\includegraphics[width=.15\textwidth]{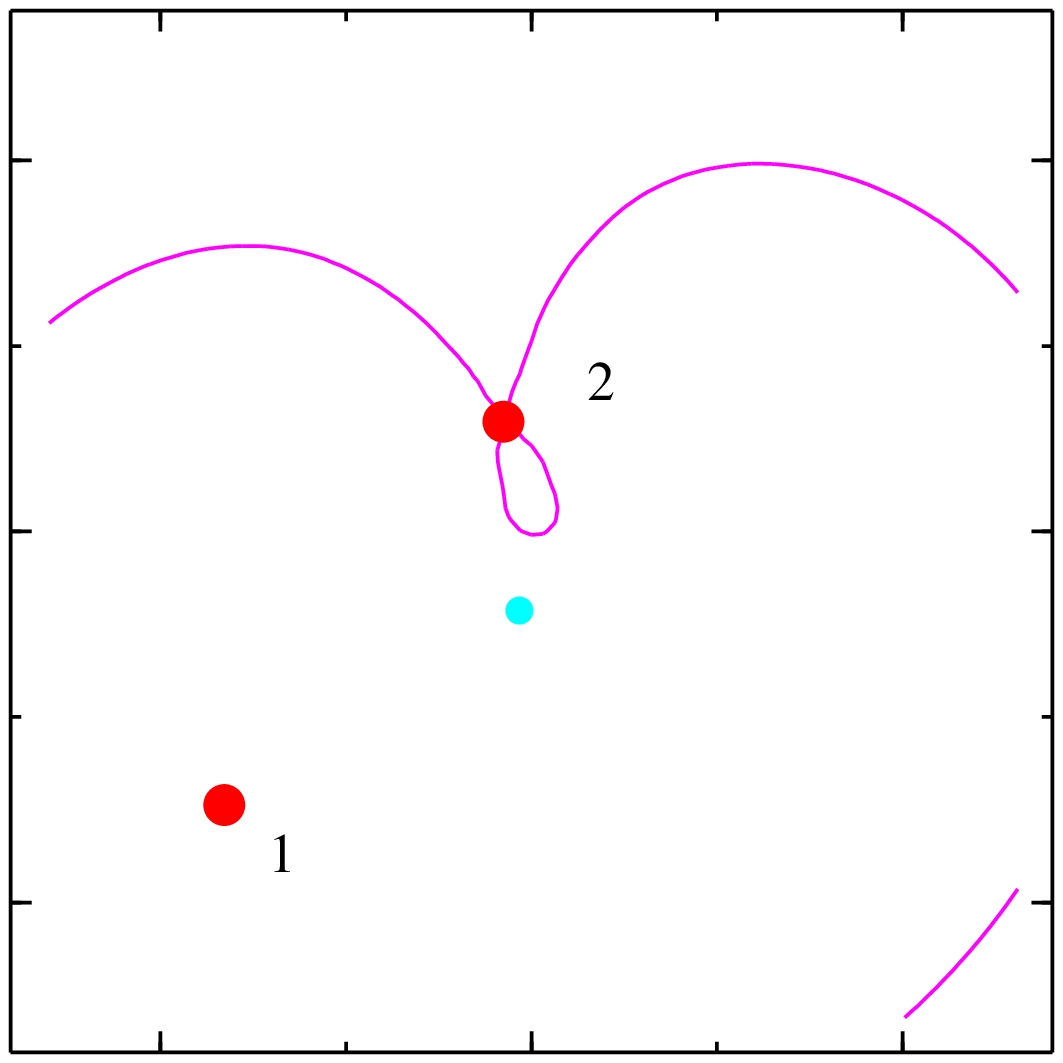}
\figskip
\includegraphics[width=.3\textwidth]{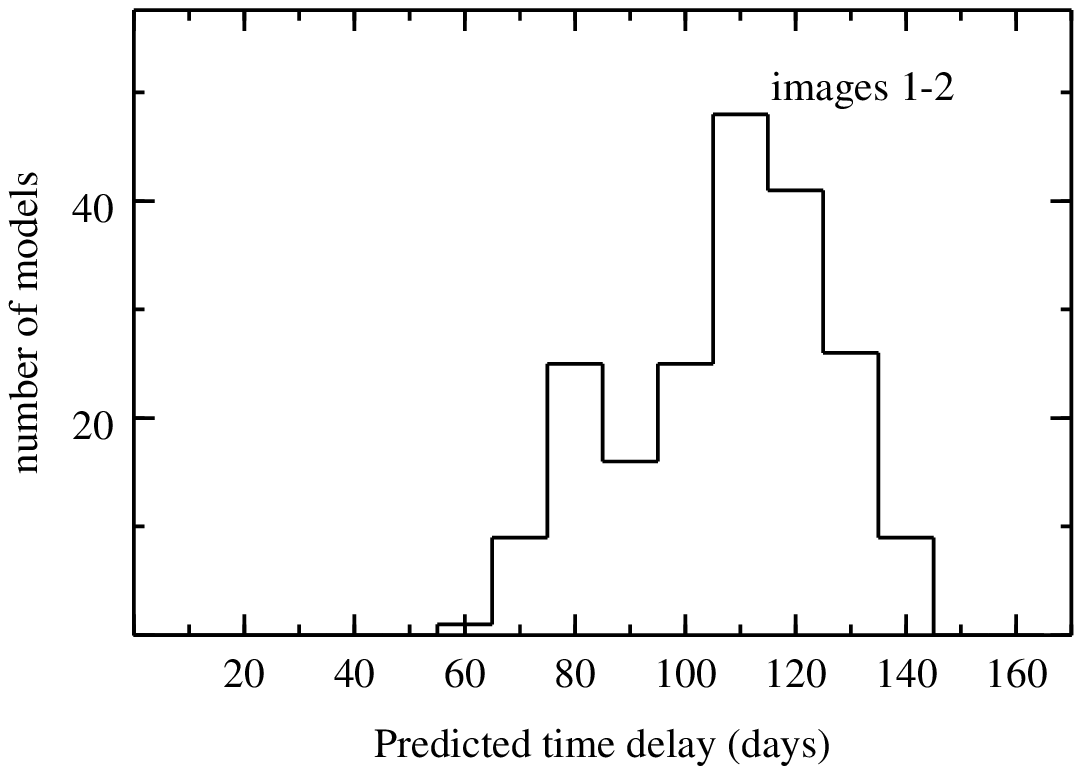}
\caption{Models of B0818+122 (inclined double). Prospects: good}
\label{plotB0818}
\end{figure}

\noindent {\bf B0818+122} [Fig.~\ref{plotB0818}] discovery: Hagen \&
Reimers (2000).  A chain of galaxies to the NE contribute a large
external shear.  The fainter image is very close to the main lensing
galaxy, and about the same brightness.  Overall, prospects appear
good.

\begin{figure}
\centering
\includegraphics[width=.15\textwidth]{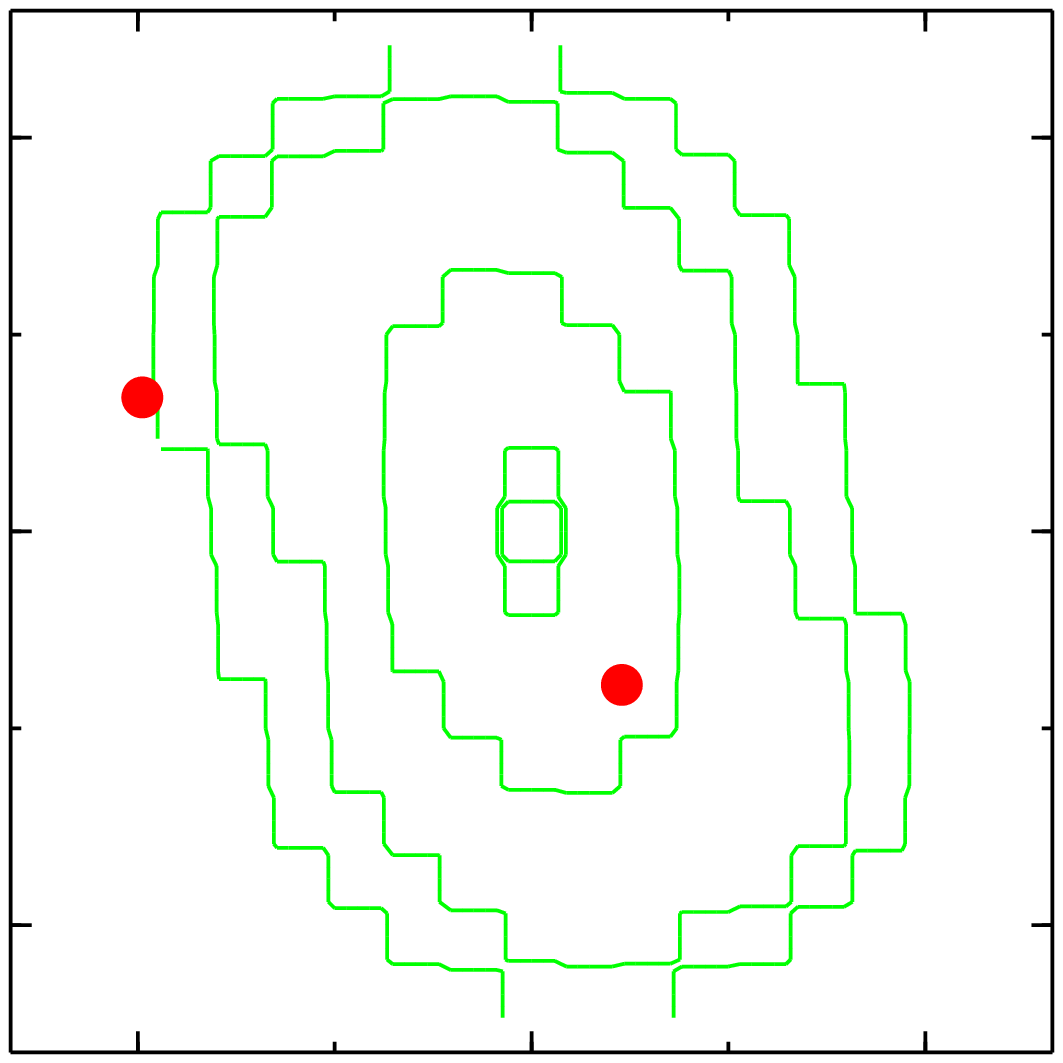}%
\includegraphics[width=.15\textwidth]{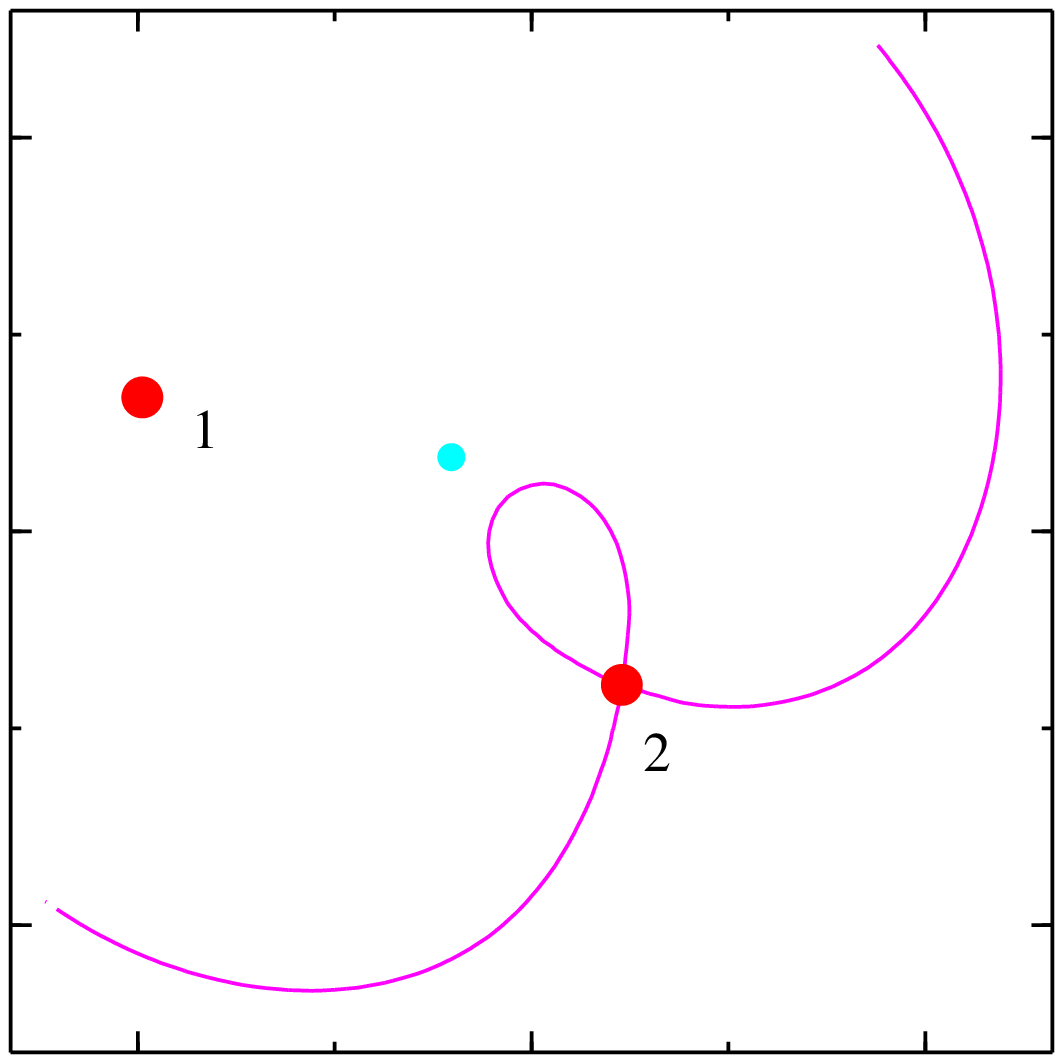}
\figskip
\includegraphics[width=.3\textwidth]{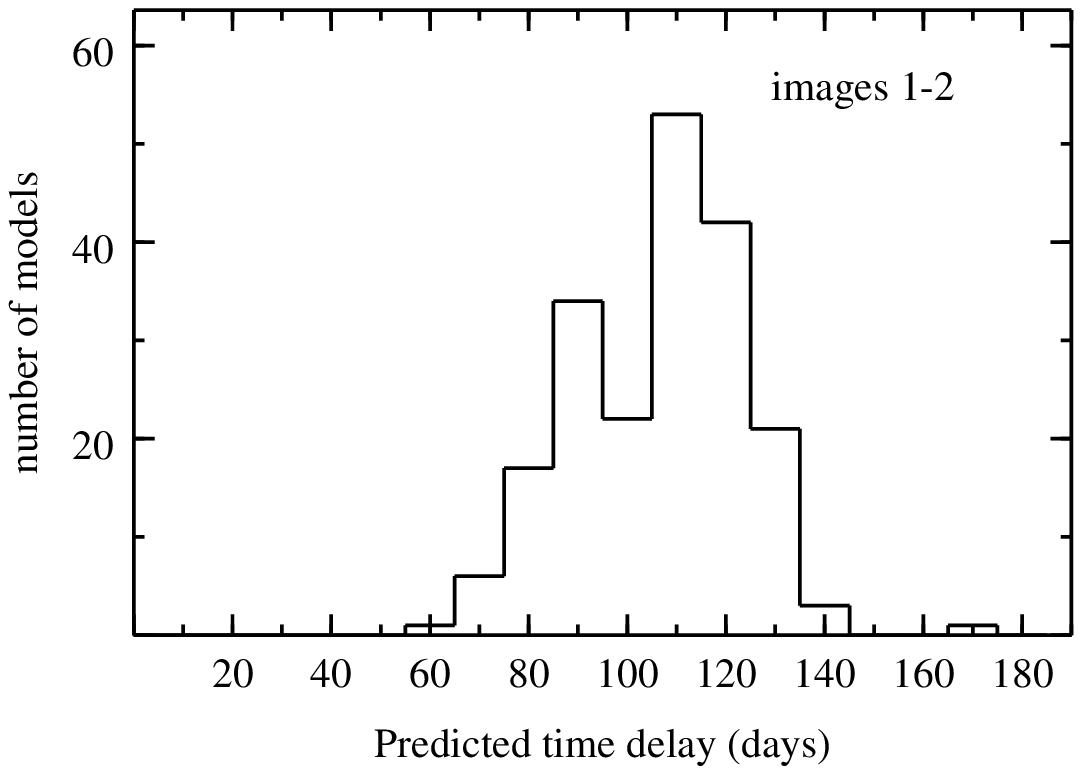}
\caption{Models of J0903+502 (inclined double). Prospects: good.}
\label{plotJ0903}
\end{figure}

\noindent {\bf J0903+502} [Fig.~\ref{plotJ0903}] discovery:
\cite{johnston03}.  There are several group galaxies in addition to
the main lensing galaxy, with one galaxy to the SW probably the major
contributor of external shear. Both quasar images are faint,
$R\sim19-20$, so monitoring is difficult with a 1\thinspace m-class
telescope.  The distribution of predicted time-delays is narrow.
Overall, we rate prospects as good.


\subsection{Long-axis quadruples}

\begin{figure}
\centering
\includegraphics[width=.15\textwidth]{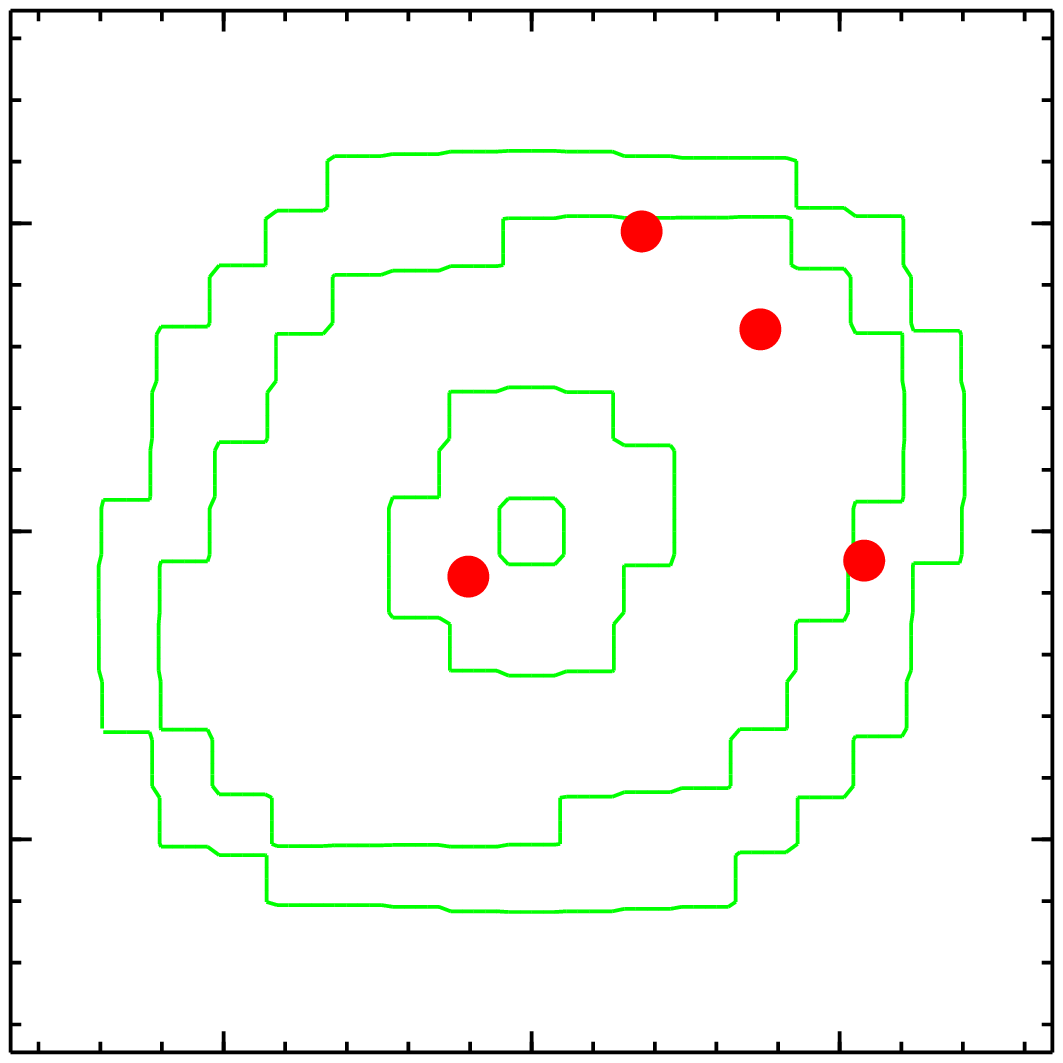}%
\includegraphics[width=.15\textwidth]{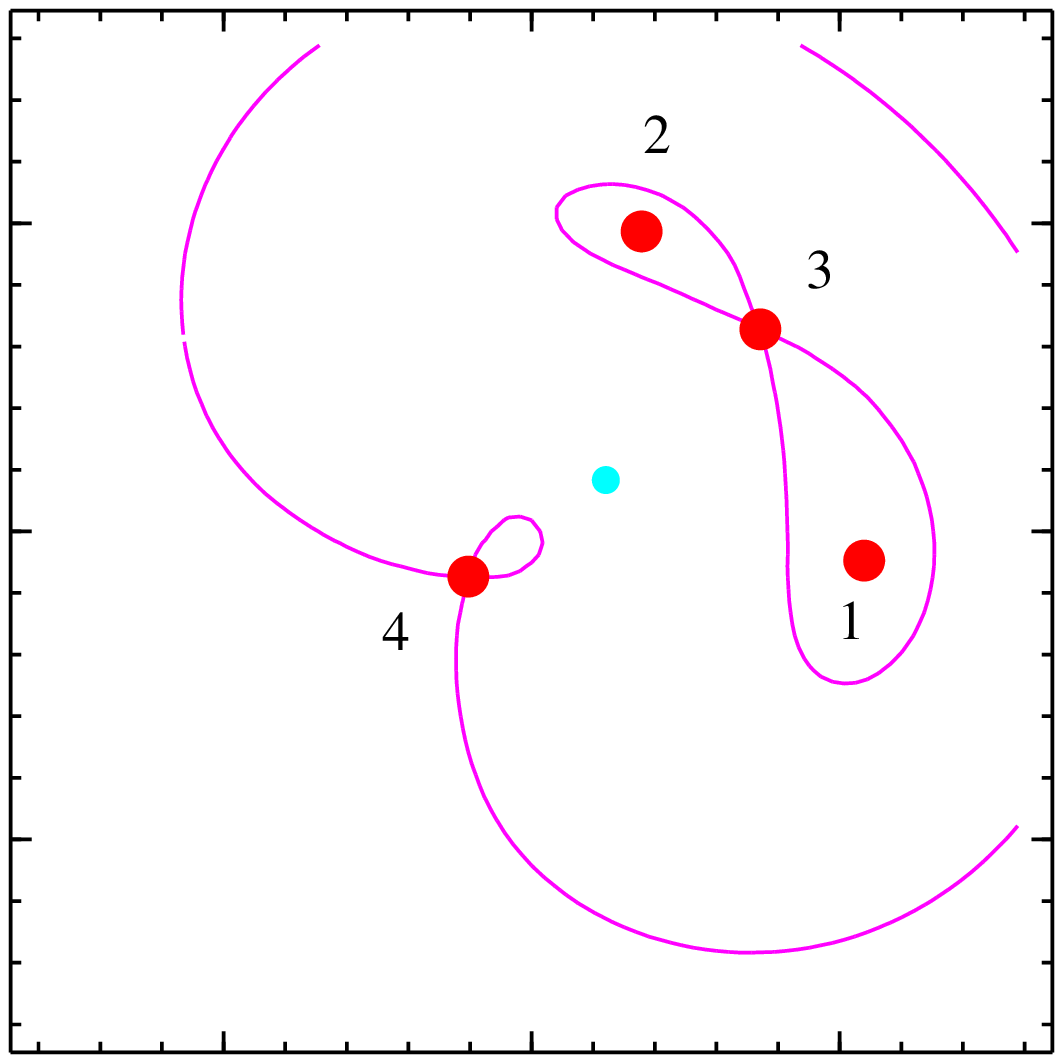}
\figskip
\includegraphics[width=.3\textwidth]{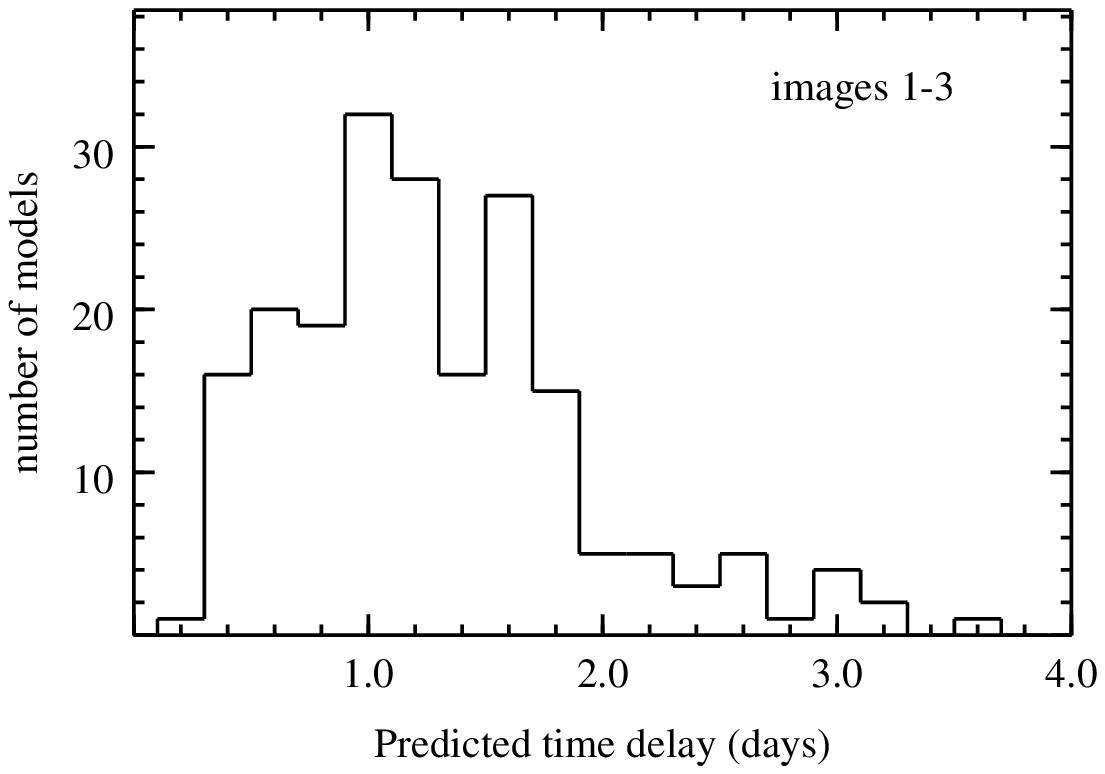}
\figskip
\includegraphics[width=.3\textwidth]{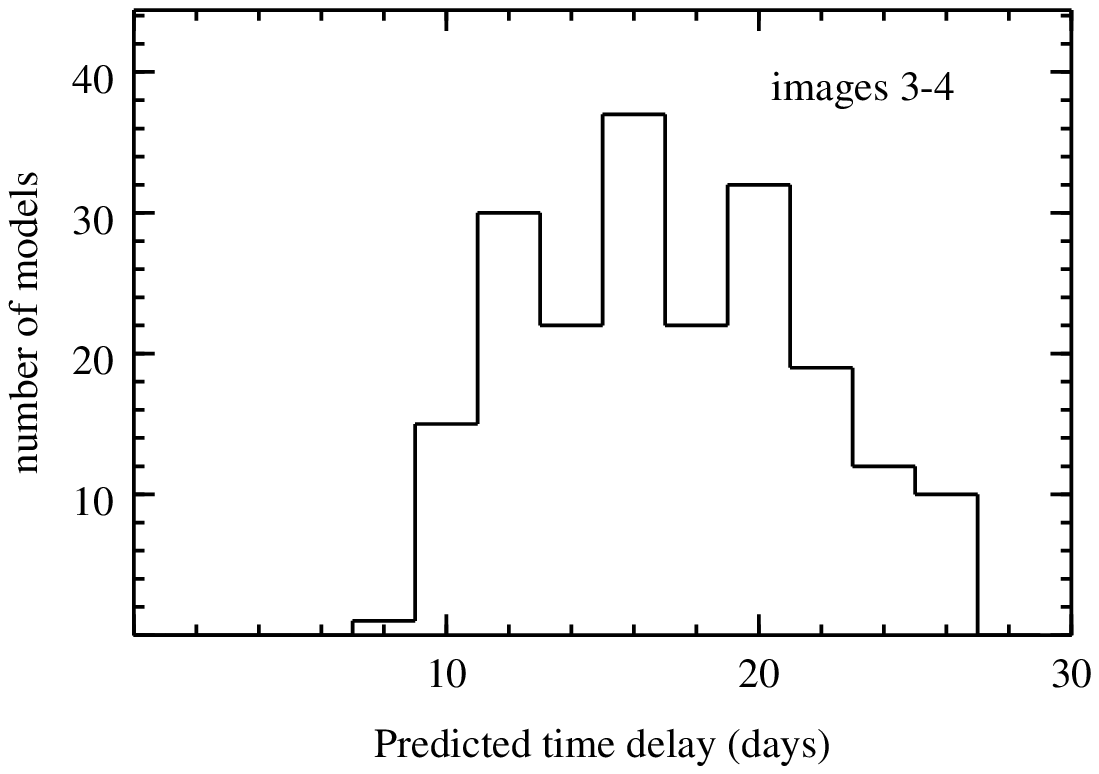}
\caption{Models of B1422+231 (long-axis quadruple). Prospects: good.}
\label{plotB1422}
\end{figure}

\noindent {\bf B1422+231} [Fig.~\ref{plotB1422}] discovery:
\cite{patnaik92}. It is a radio emitter with extremely accurate image
positions.  There is evidence of variability.  Strong external shear
comes from a galaxy group to the SE.  Time-delays between the close
triplet of images may be too short to measure in the optical, but the
delay to the fourth image can be expected to be useful.  The lensing
galaxy is comparable in brightness to the faint fourth image, which
complicates the photometry.  Overall, prospects appear good.

\begin{figure}
\centering
\includegraphics[width=.15\textwidth]{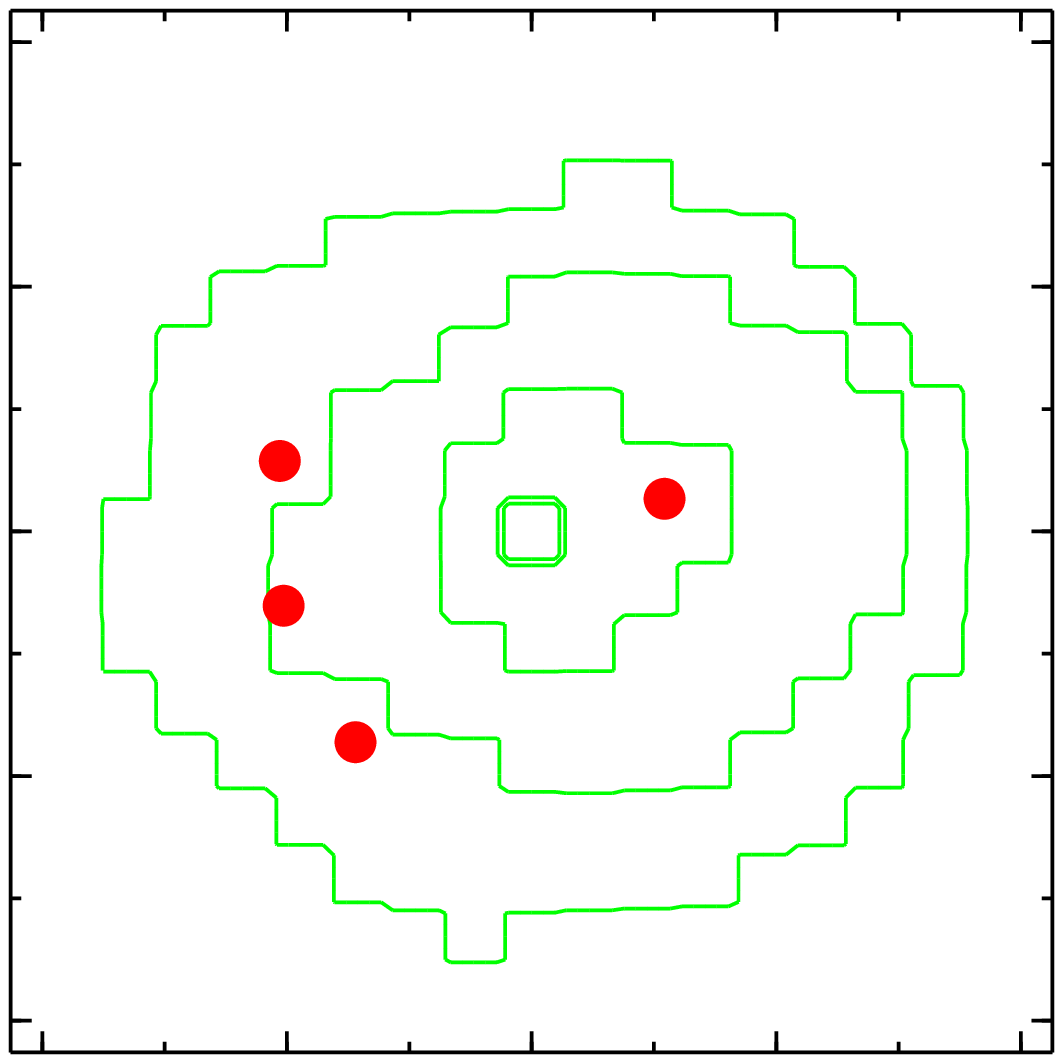}%
\includegraphics[width=.15\textwidth]{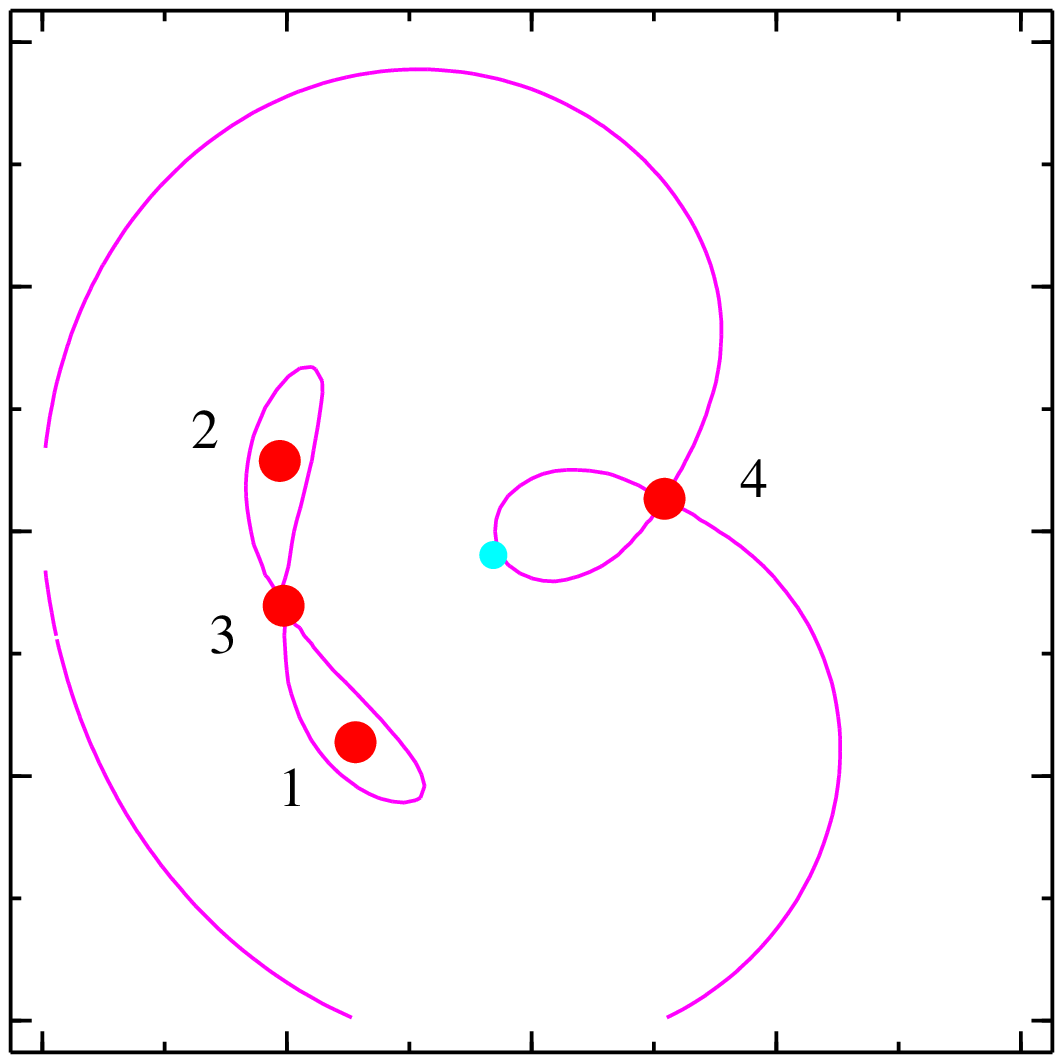}
\figskip
\includegraphics[width=.3\textwidth]{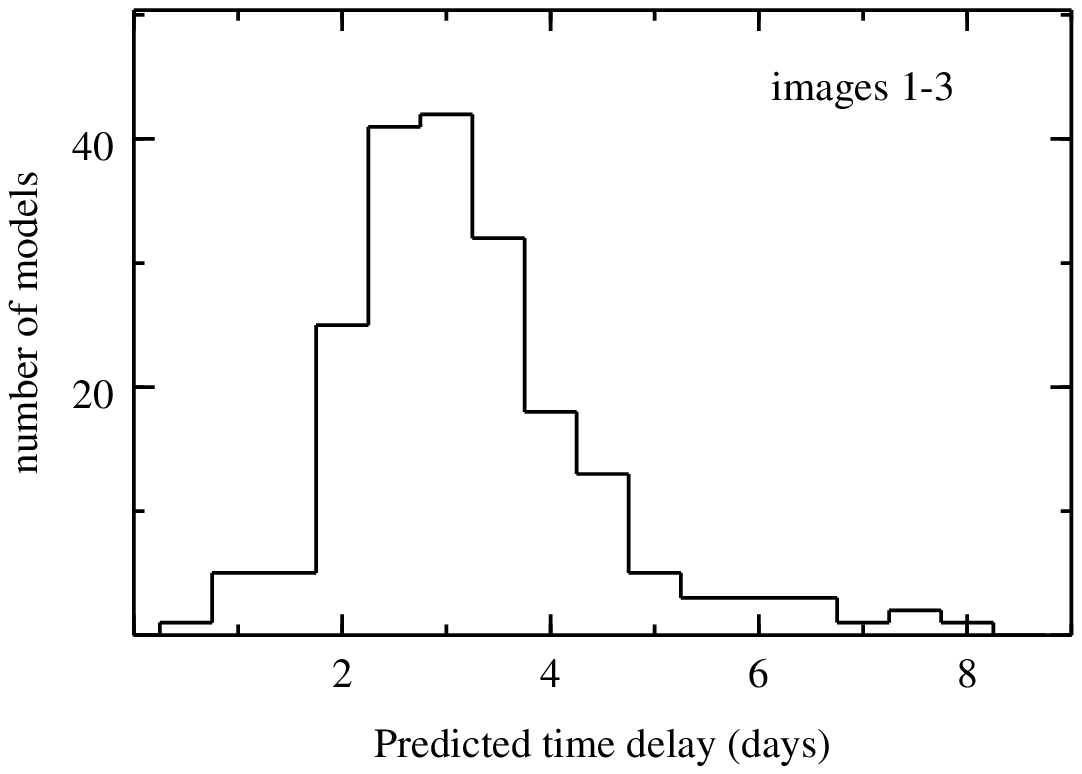}
\figskip
\includegraphics[width=.3\textwidth]{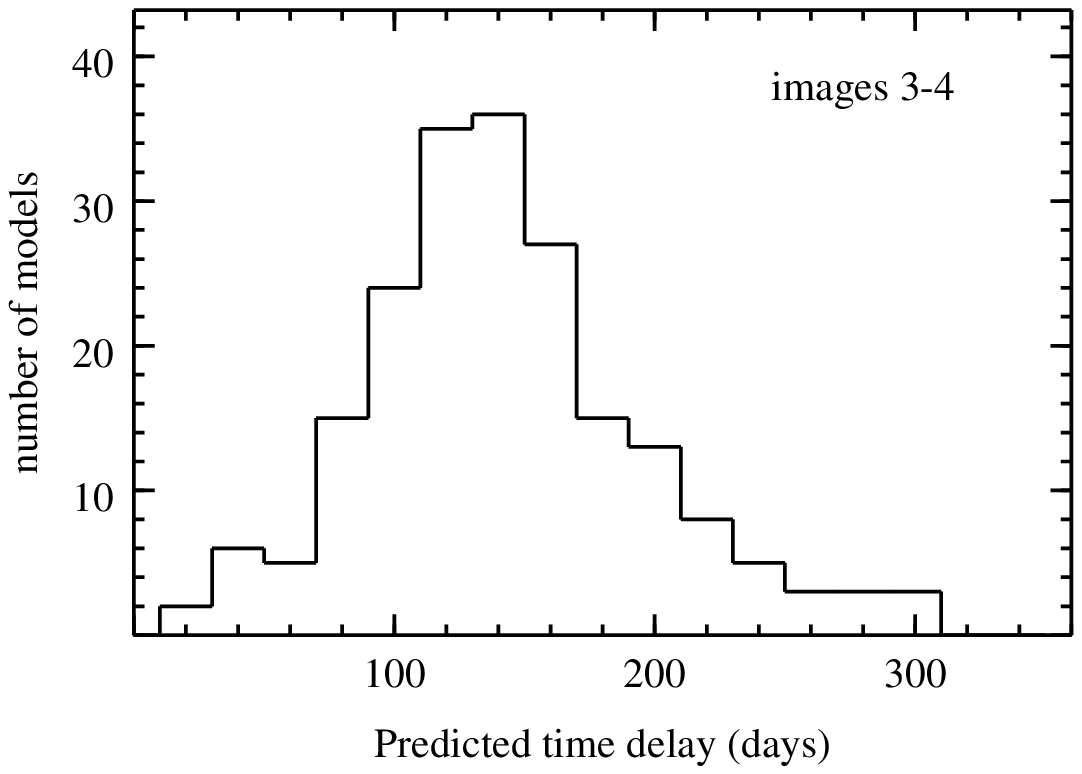}
\caption{Models of J1131--123 (long-axis quadruple). Prospects: excellent.}
\label{plotJ1131}
\end{figure}

\noindent {\bf J1131--123} [Fig.~\ref{plotJ1131}] discovery:
\cite{sluse03}.  It is a quadruple with large separation: $3.69''$.  It
is very like a larger sibling of B1422+231. Morphology and models
indicate significant external shear from mass to the WNW or ESE. There
is evidence for intrinsic variability.  Structures in the Einstein ring
are likely to offer additional model constraints, though they also
contaminate the flux from the faint fourth image.  Overall, prospects
appear excellent.


\subsection{Inclined quadruples}

\begin{figure}
\centering
\includegraphics[width=.15\textwidth]{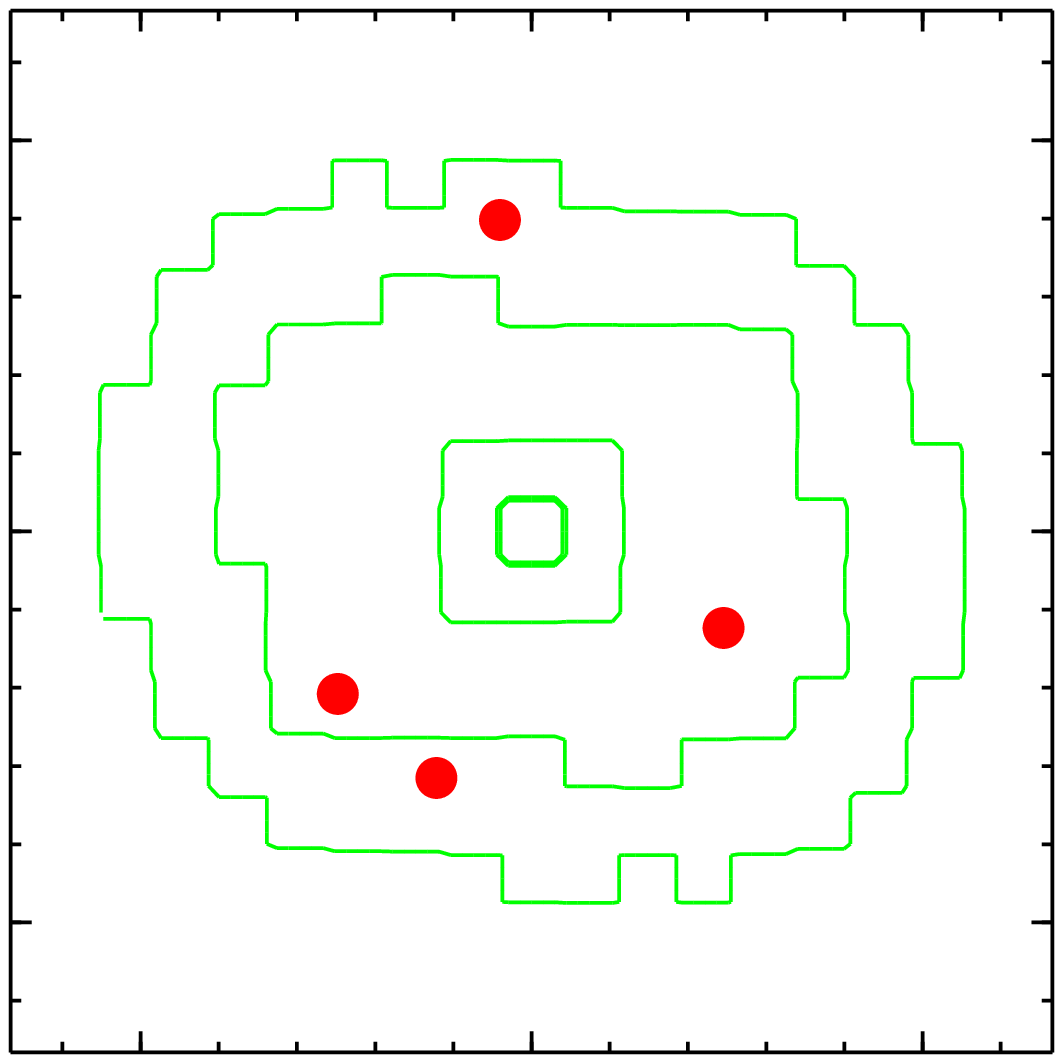}%
\includegraphics[width=.15\textwidth]{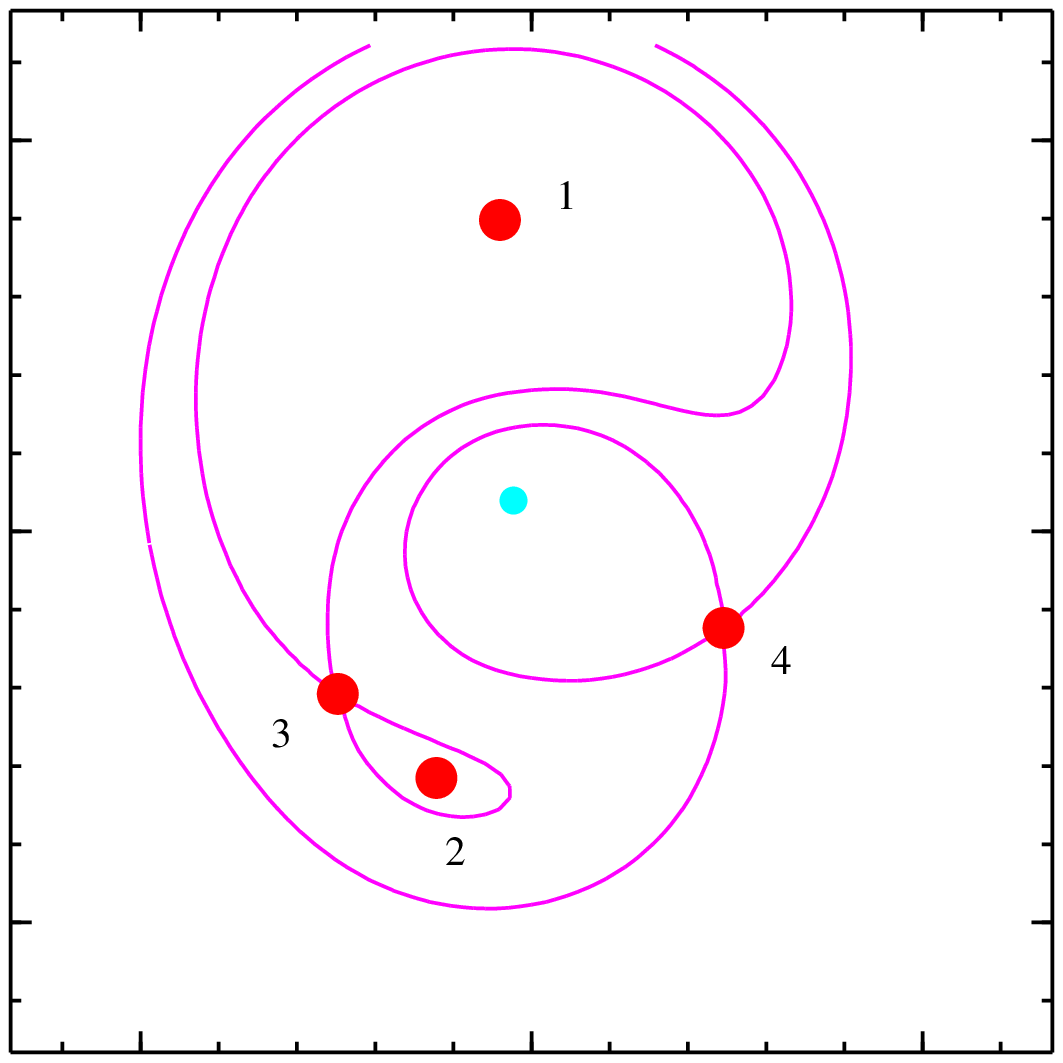}
\figskip
\includegraphics[width=.3\textwidth]{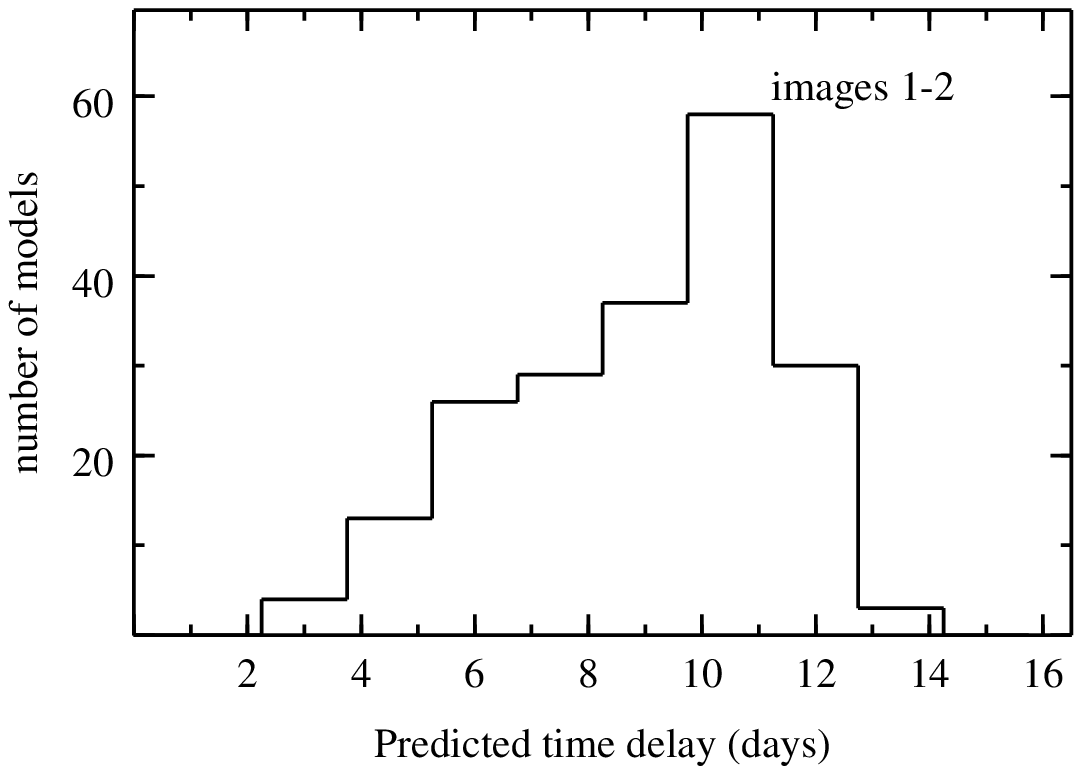}
\figskip
\includegraphics[width=.3\textwidth]{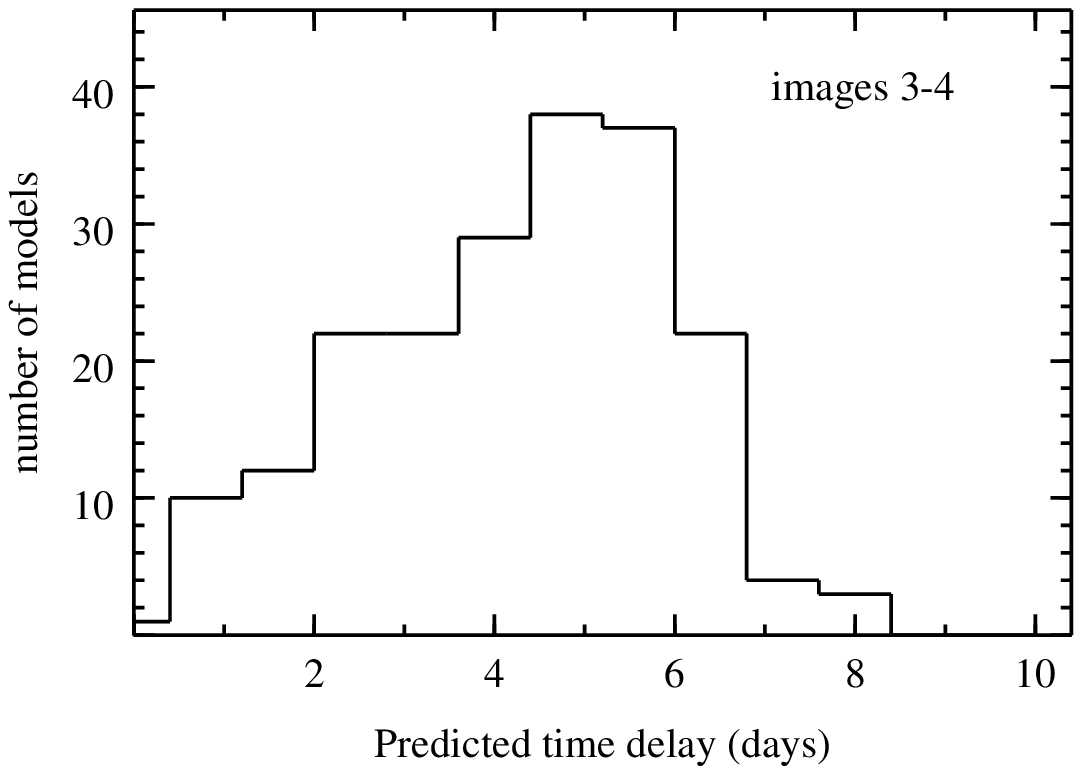}
\caption{Models of J2026--453 (inclined quadruple). Prospects:
good.}
\label{plotJ2026}
\end{figure}

\begin{figure}
\centering
\includegraphics[width=.15\textwidth]{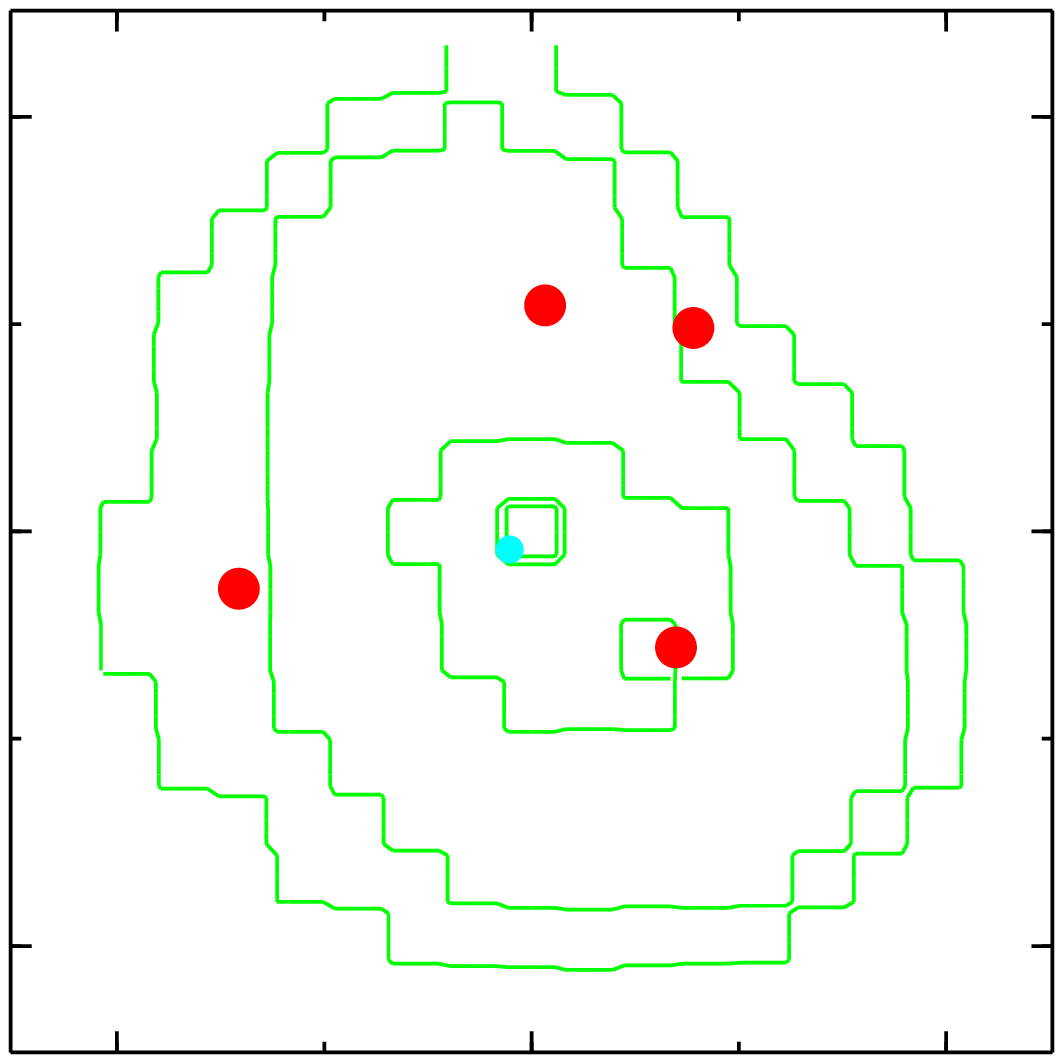}%
\includegraphics[width=.15\textwidth]{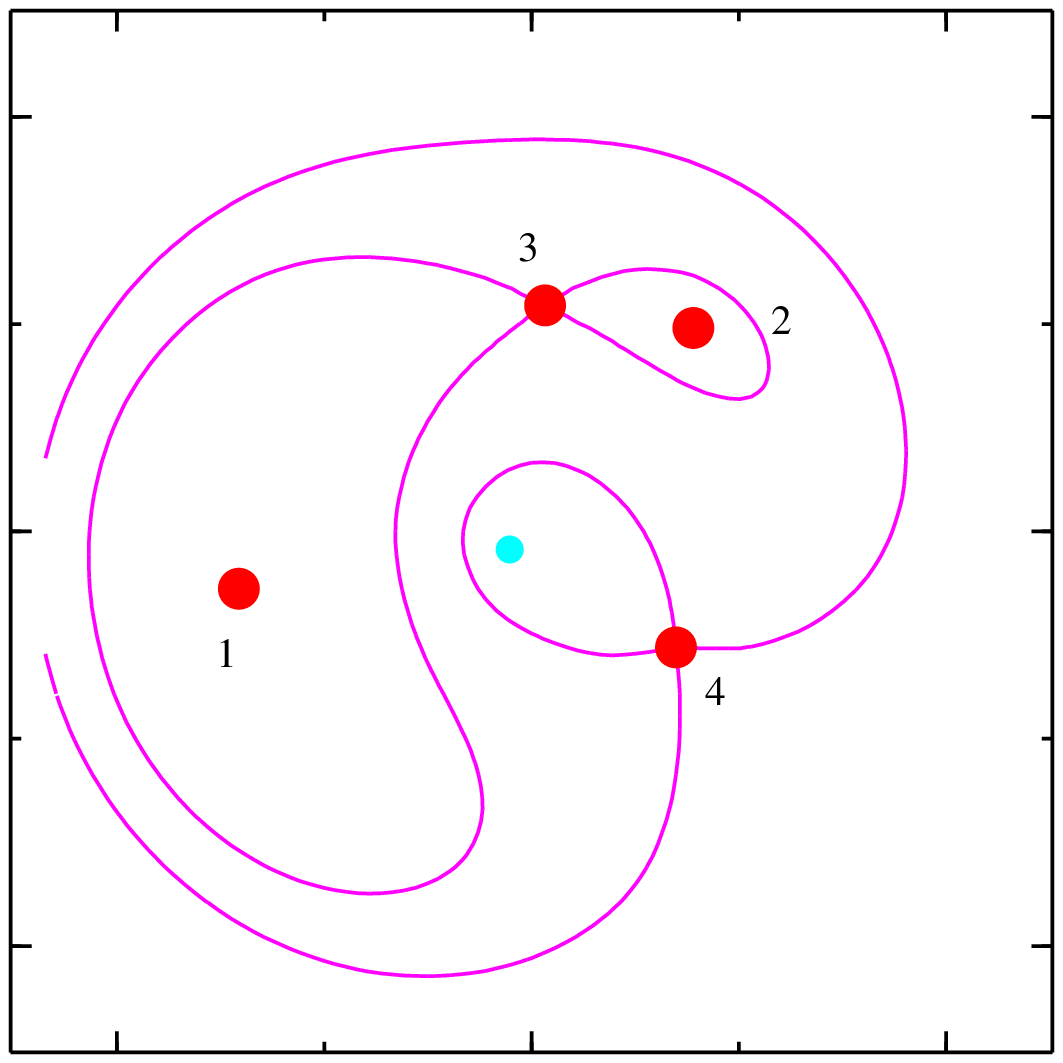}
\figskip
\includegraphics[width=.3\textwidth]{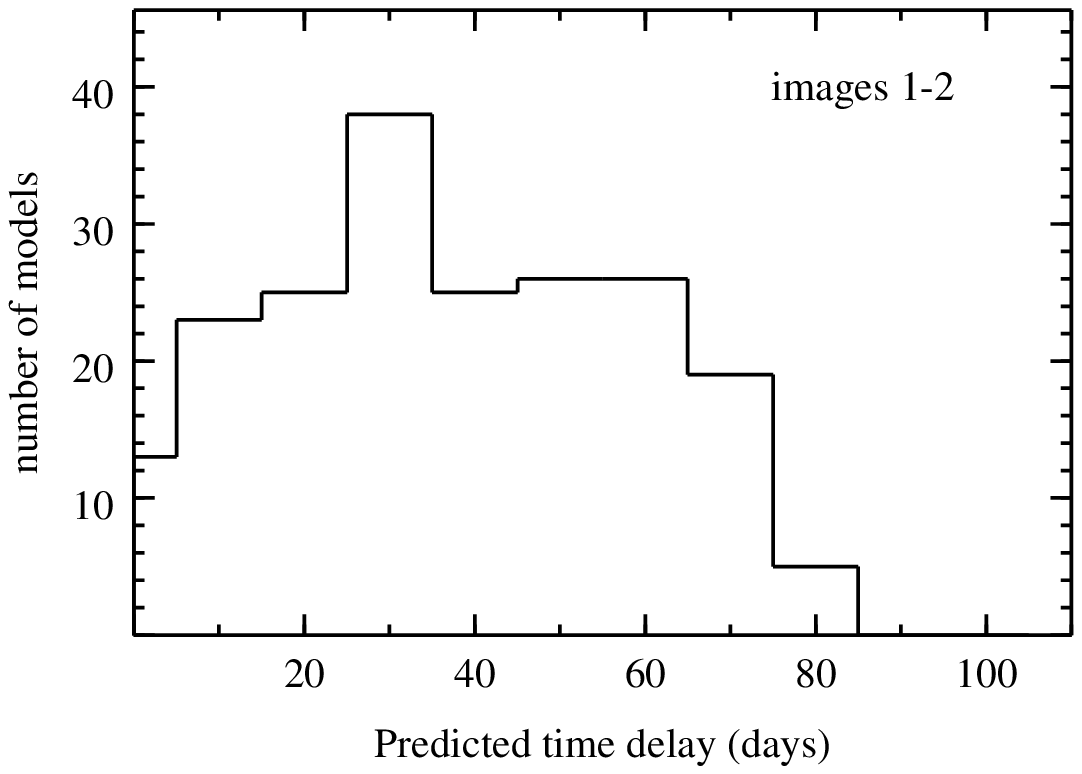}
\figskip
\includegraphics[width=.3\textwidth]{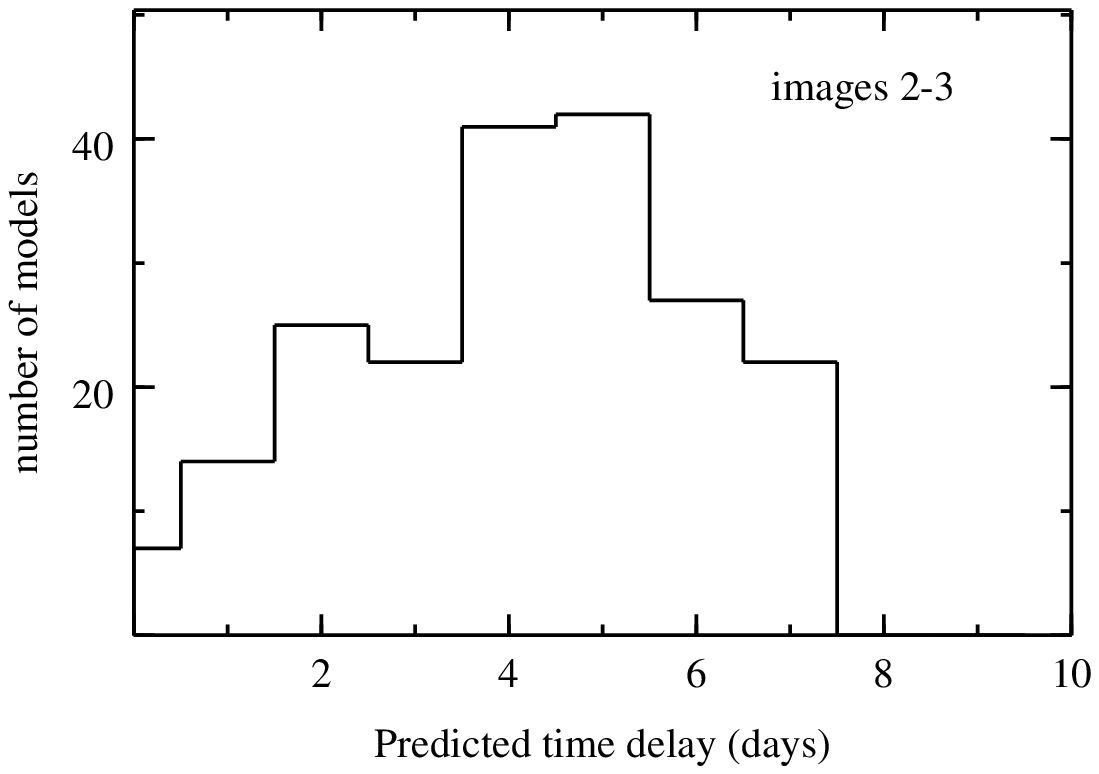}
\figskip
\includegraphics[width=.3\textwidth]{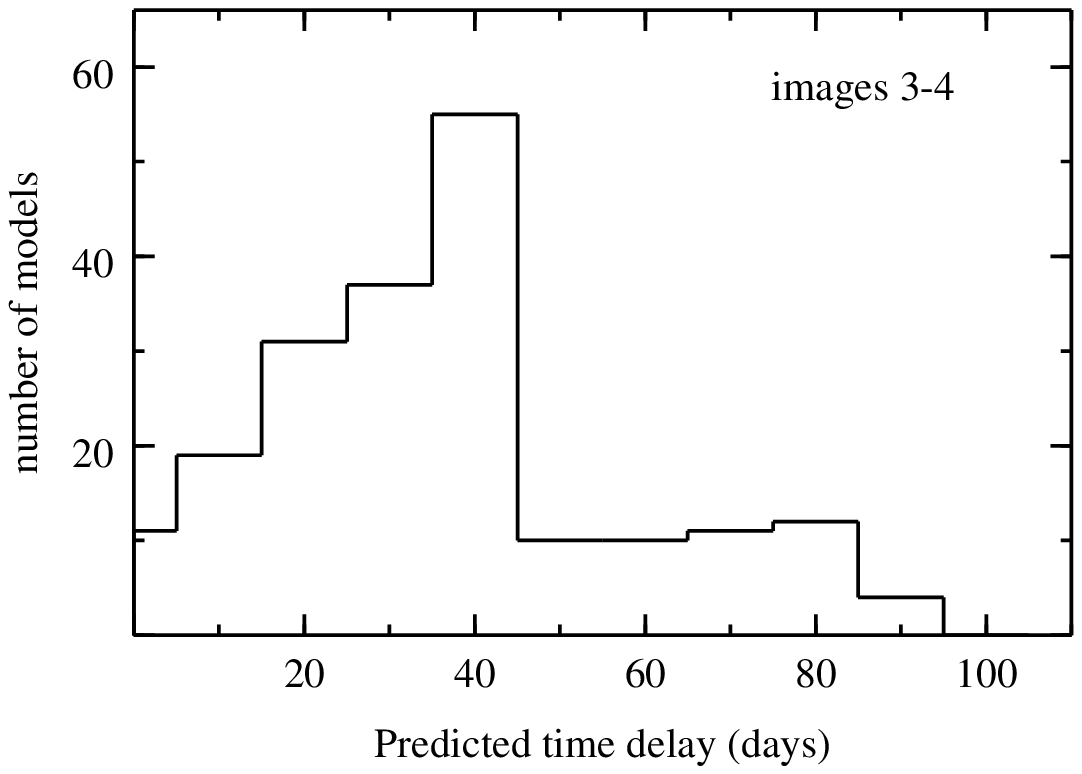}
\caption{Models of J2033--472 (inclined quadruple). Prospects: excellent.}
\label{plotJ2033}
\end{figure}

\noindent {\bf J2026--453} [Fig.~\ref{plotJ2026}] and {\bf J2033--472}
[Fig.~\ref{plotJ2033}] discovery: \cite{morgan04}. In J2026-453,
morphology and models indicate external shear from mass to the E or W.
This system so far lacks a $\zl$; we assumed 0.5, which is plausible
given the colours of the galaxy.  The morphology of J2033--472
suggests an asymmetric lens, and accordingly we have considered
asymmetric models.  We rate time-delays prospects as good for J2026
and excellent for J2033.

\begin{figure}
\centering
\includegraphics[width=.15\textwidth]{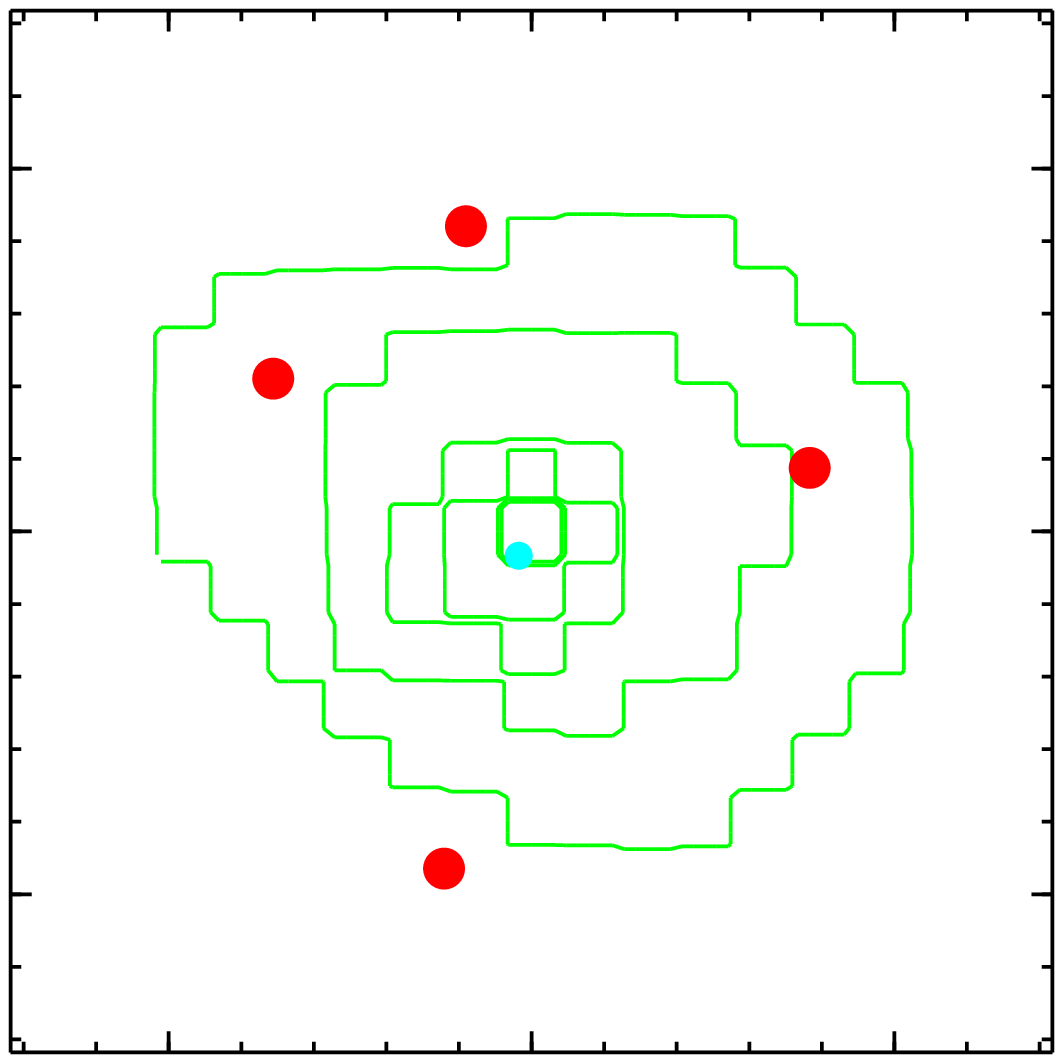}%
\includegraphics[width=.15\textwidth]{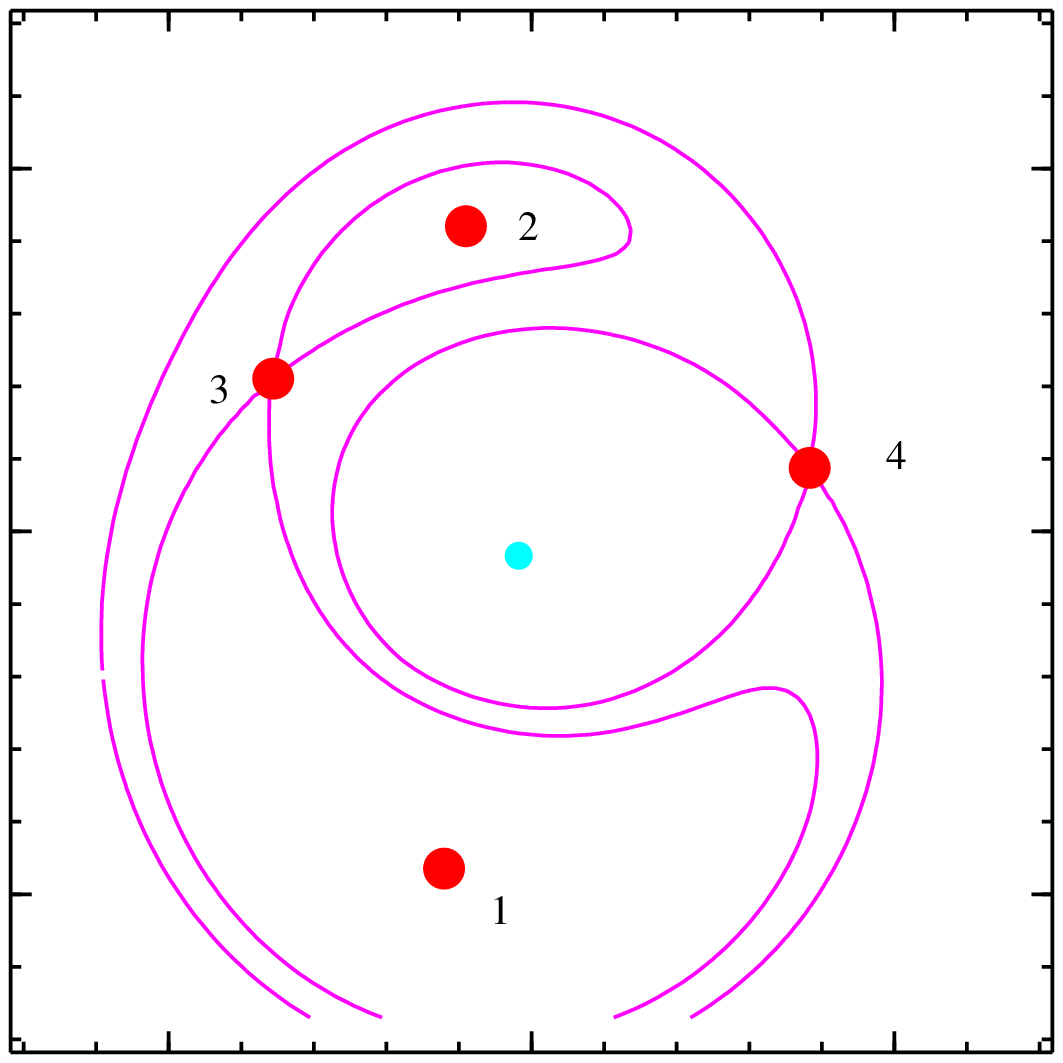}
\figskip
\includegraphics[width=.3\textwidth]{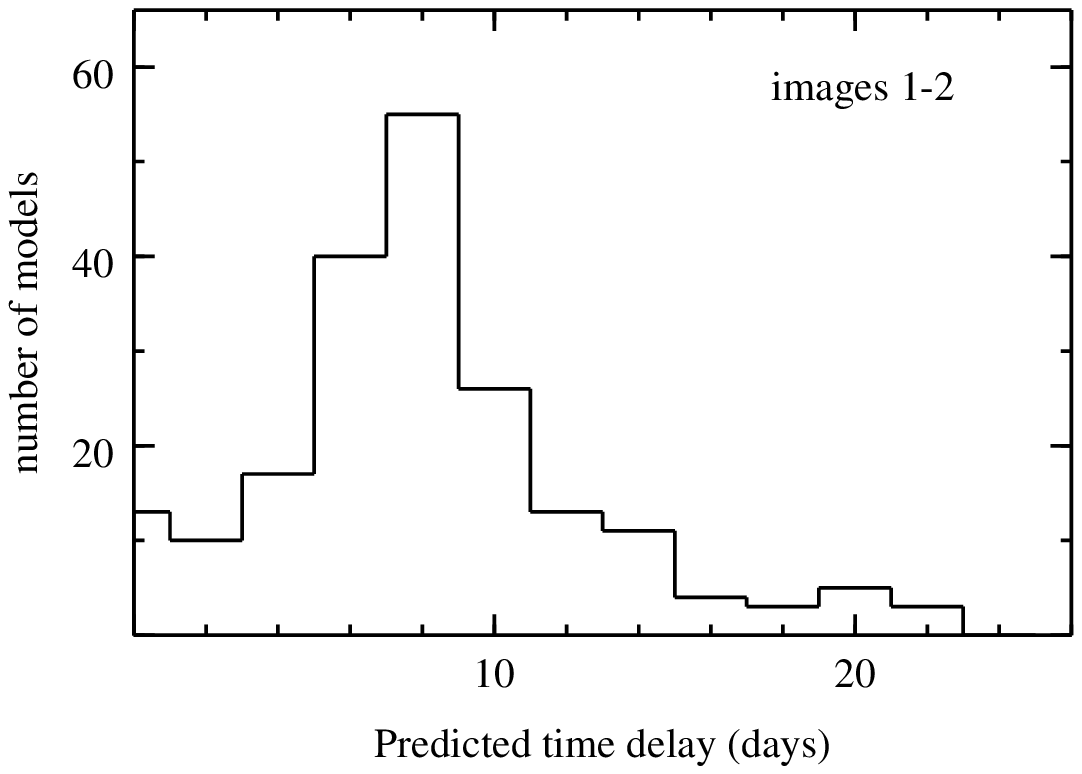}
\figskip
\includegraphics[width=.3\textwidth]{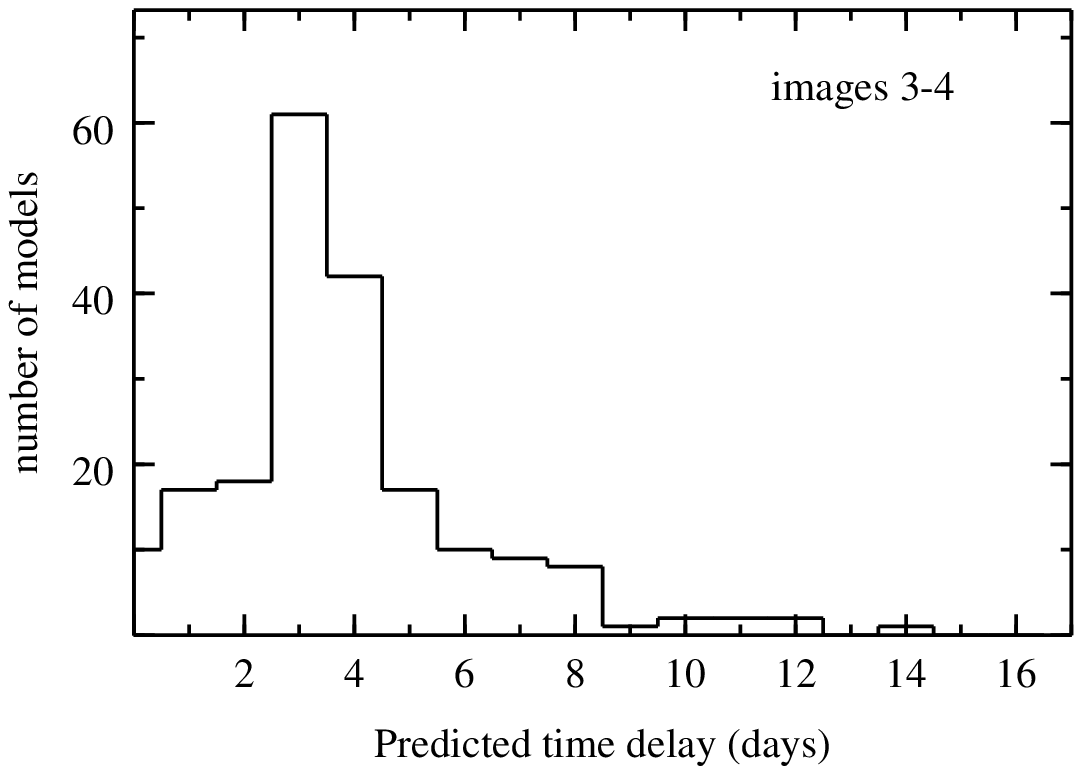}
\caption{Models of J0924+021 (inclined quadruple). Prospects:
currently unpromising.}
\label{plotJ0924}
\end{figure}

\noindent {\bf J0924+021} [Fig.~\ref{plotJ0924}] discovery:
\cite{inada03}.  This is a complex and evidently asymmetric lens, but
can be well-modelled and leads to relatively tight predictions for two
of the time delays.  However, image 3 is very faint, which
\cite{keeton05} argue is the result of microlensing.  This greatly
complicates the measurements of time delays, so we current rate this
lens as an unpromising time-delay prospect.


\section{Predicted precision for the Hubble time} \label{predacc}

In the previous section, after considering detailed models as well as
observational circumstances of all 15 lenses, we concluded that 5
systems are excellent candidates for time-delay monitoring (J1650+425,
J2033--472, B0909+532, J1335+011, J1131--123), and 6 are good
candidates (J1355--225, J0903+502, B0818+122, B1009--025, B1422+231,
J2026--453).  We now ask how accurately $H_0^{-1}$ can be inferred if
the 5 excellent candidates, or if the 11 excellent or good candidates,
have their time delays measured accurate to 1\thinspace d.  We do not
expect that either scenario will be what transpires in the future.  We
expect that some of these 11 lenses will yield accurate time delays
over the next 2--3 years, while some existing time-delay measurements
are refined.  But the 5-lens and 11-lens cases are reasonable
surrogates for a future set of available measurements.

\begin{figure}
\centering
\includegraphics[width=.3\textwidth]{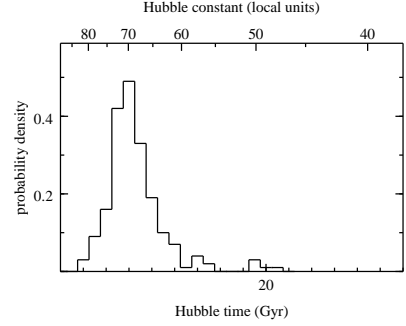}
\caption{Hubble constant/time recovered from simulated time delays of
the 5 excellent candidates.  The confidence intervals (read off by
sorting the unbinned values) are $14.0^{+1.2}_{-0.7}$Gyr at 68\%
and $14.0^{+2.9}_{-1.3}$Gyr at 90\%.}
\label{sim-excel5}
\end{figure}

Fig.~\ref{sim-excel5} shows the recovered $H_0^{-1}$ from simulated time
delays of the 5 excellent candidates.  For each lens we took a random
model (from the ensemble of 200), read off its time delays rounded to
the nearest day, and then took them as simulated time delays. Any model
delays of $\leq1$d we treated as unmeasured.  Using {\em PixeLens,} we
then modelled the 5 lenses simultaneously from the actual image
positions and these simulated time delays.  The model-ensemble had 200
members, each member consisting of models for all 5 lenses sharing a
common $H_0^{-1}$ (\cite{sw04}).  Fig.~\ref{sim-excel5} shows the resulting 200
values of $H_0^{-1}$ after binning.  We see that the original input
value 14~Gyr is recovered with $<10\%$ uncertainty at 68\% confidence,
and no discernable bias.  Also, the uncertainties are asymmetric.

\begin{figure}[t!]
\centering
\includegraphics[width=.3\textwidth]{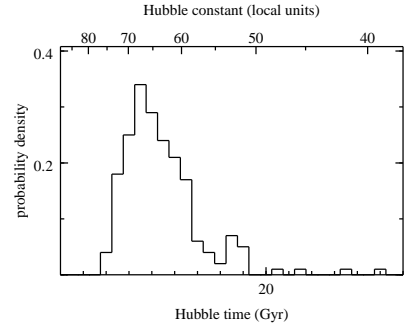}
\caption{Hubble constant/time from simulated time delays of the 6 good
candidates.  The confidence intervals are $15.0^{+1.3}_{-1.0}$~Gyr at
68\% and $15.0^{+3.7}_{-1.5}$~Gyr at 90\%.}
\label{sim-good6}
\end{figure}

Fig.~\ref{sim-good6} shows the result of a similar exercise using the
6 good candidates.  The uncertainties are somewhat larger than in
Fig.~\ref{sim-excel5} and similarly asymmetric.  Also, there is a
bias, in the sense that the median value is $15.0$~Gyr rather than
$14.0$~Gyr; but the bias is in the 68\% confidence range and hence
not significant.

\begin{figure}
\centering
\includegraphics[width=.3\textwidth]{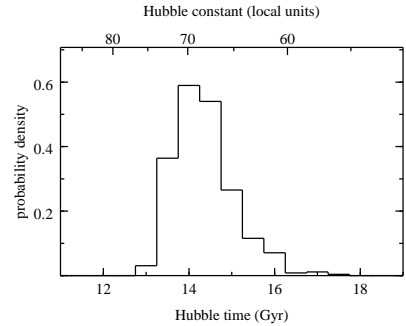}
\caption{Hubble constant/time from simulated time delays of the 11
excellent or good lenses.  This histogram is simply the product of the
two previous ones.  From a narrower binning, we read off the
confidence intervals $14.2^{+0.7}_{-0.7}$Gyr at 68\% and
$14.4^{+1.9}_{-0.8}$Gyr at 90\%.}
\label{sim-best11}
\end{figure}

The 5-lens and 6-lens ensembles just described are independent.  Hence
we can simply multiply their histograms.  Fig.~\ref{sim-best11} shows
the result.  We recover $H_0^{-1}$ with an uncertainty of about 5\%,
at 68\% confidence.

These results show that the Hubble time can be recovered to 5\%
precision even if we allow for a large diversity in possible mass
distributions (or prior).  But there is a caveat, which needs to be
addressed before a claim of 5\% accuracy (rather than precision) can
be made.  Currently, a mass distribution that satisfies the lensing
constraints is either allowed by the prior as a plausible galaxy lens,
or rejected; there is no weighting in the prior.  Properly, the prior
should weight mass models according to their abundance in the real
world of galaxies.  Lack of weighting will introduce a bias.  (This
prior-induced bias is different from the small statistical bias seen
in Fig.~\ref{sim-good6}.)  The blind tests in Williams \& Saha (2000)
would have detected biases if they were around 20\% or more.  But
prior-induced biases at the 5\% level remain untested for.  Finding
them and then eliminating them through a weighted prior could be done
by calibrating against galaxy-formation models, and is an essential
theoretical program needed to complement the observations.


\section{Discussion}

In this paper we do three things: first, we introduce a simple rough
predictor for time delays $\Tastrom$, second, we make model predictions
for 23 time delays covering 14 lenses, and finally we estimate the
precision in the Hubble time inferred from simulated data on the 11 best
lenses.  The main conclusion is that no single lens can usefully
constrain $H_0$, but time delays accurate to $\simeq1$d on $>10$ lenses
can yield $H_0$ accurate to 5\%.

In the histograms in Figs.~\ref{plotJ1155}--\ref{plotJ0924}, typically
90\% of the area ranges over a factor of two in time delays.  Hence, a
monitoring program can have 90\% confidence in succeeding --- provided
the quasar is sufficiently variable --- if the sampling allows for the
appropriate 90\%-range of possible time delays.  There is no single
characteristic shape for the histograms, but the pattern of a low-end
tail and a high-end cliff is common.  The large uncertainty in the
predicted time delays reflects the large variety of mass models models
that can reproduce the observed image positions in any given lens.
The prior we have for deciding what is an allowable mass model for a
galaxy is very conservative, so our models have more variety than
reality.  But not very much more --- if all real galaxy lenses
belonged to some known parametrization, then error-bars on $H_0$ from
fitting such parametric models to observed time-delay systems would
overlap, but as is evident from Fig.~12 in \cite{cou03}, those
error-bars do not overlap.

Actually, the basic results about predicted time delays are already
present in the summaries Figs.~\ref{plotobs} and \ref{plocand}.  To
interpret these figures, recall that $\Tastrom$ makes a preliminary
prediction for the time-delay, while the deviation from the oblique
line depends on the details of mass distribution and lens morphology.
From Fig.~\ref{plotobs} we see that $\Tastrom$ gets us to within a
factor of two of the observed values.  Now, the error bars ---note that
the error bars in tables and figures are 68\% confidence--- in
Fig.~\ref{plocand} are generally less than a factor of two; thus
detailed modelling does provide a better prediction than $\Tastrom$
alone, but not dramatically better.  We can also see that several of
the error bars in Fig.~\ref{plocand} are shorter on top, thus
indicating a low-end tail and a high-end cliff.

The image morphology of a lens is correlated with the uncertainty in
the time delays, especially in quadruples.  Core quadruples
generically have short time delays and are unlikely to be useful for
time delays; the case of B0435-122 is illustrative.  Long- and
short-axis quadruples, having three images close together, are likely
to have only one measurable time delay.  Inclined quadruples are the
most promising, since they usually have two time delays in the
measurable range, and sometimes three.  Among doubles, inclined
systems tend to be somewhat better constrained than axial systems.
Significant asymmetry in the lens is a disadvantage, but in
compensation, asymmetry increases the chance of having three
measurable time delays.  Thus B1608+656 is an asymmetric inclined
quadruple with three measured time delays; J2033--472 may prove to be
another, and is among our excellent candidates.  Surprisingly, a large
external shear appears to reduce uncertainties in the time delay.
This is particularly noticeable in the inclineddoubles J1650+425,
B0909+532, and J0903+502, and the short-axis quads B1422+231 and
J1131--123. The reason is not clear; it may be that since external
shear reduces amount of mass needed in the main lens to produce
multiple images, it reduces the available model-space.

Finally, the results from combining several lenses are very encouraging.
Assuming time delays accurate to 1\thinspace d we find that the
model-dependent uncertainty in $H_0^{-1}$ reduces to less than 10\% on
combining the 5 best lenses, and about 5\% on combining the best 11
lenses.  The uncertainties are asymmetric, with the lower limit on the
Hubble time being tighter than the upper limit.  More work needs to be
done on the model prior before we can truly attain 5\% accuracy, but
meanwhile our results help provide both motivation and observing
strategies for accurate time-delay measurements.

\begin{acknowledgements}
The authors thank Dr. Steve Warren for very helpful discussions.
COSMOGRAIL is supported by the Swiss National Science Foundation (SNSF).
\end{acknowledgements}


\begin{thebibliography}{}

\bibitem[Barkana (1997)]{barkana97}
Barkana, R. 1997, \apj, 489, 21

\bibitem[Bernstein \& Fischer 1999]{bernstein99}
Berstein, G., Fischer, P. 1999, \aj, 118, 14

\bibitem[Biggs et al.\ (1999)]{biggs99}
Biggs, A. D., Browne, I. W. A., Helbig, et al.\ 1999, \mnras, 304, 339

\bibitem[Binney et al.\ (1991)]{binney91}
Binney, J., Gerhard, O. E., Stark, et al.\ 1991, \mnras, 252, 210 

\bibitem[Burud et al.\ (2000)]{burud00}
Burud, I., Hjorth, J., Jaunsen, A. O., et al.\ 2000, \apj, 544, 117

\bibitem[Burud et al.\ (2002a)]{burud02a}
Burud, I., Courbin, F., Magain, P., et al.\ 2002, A\&A, 383, 71

\bibitem[Burud et al.\ (2002b)]{burud02b}
Burud, I., Hjorth, J., Courbin, F., et al., 2002, A\&A, 391, 481

\bibitem[Cohen et al.\ (2000)]{cohen00}
Cohen, A. S., Hewitt, J. N., Moore, C. B. \& Haarsma, D. B., 2000
\apj, 545, 578

\bibitem[Courbin et al. \ 2002]{cou02}
Courbin, F., Meylan, G., Kneib, J.-P., Lidman, C. \ 2002, ApJ 575, L95

\bibitem[Courbin (2003)]{cou03}
Courbin, F. 2003, {\tt astro-ph/0304497}

\bibitem[Eigenbrod et al. (2005)]{eigen05} Eigenbrod, A., Courbin, F.,
  Vuissoz, C., et al. \ 2005, A\&A 436, 25

\bibitem[Fassnacht et al.\ (2002)]{fassnacht02}
Fassnacht, C. D., Xanthopoulos, E., Koopmans L. V. E. \& Rusin D. 2002
\apj, 581, 823

\bibitem[Freedman et al.\ 2001]{freedman01}
Freedman, W. L., Madore, B. F., Gibson, B. K. et al.\ 2001, \apj, 553, 47

\bibitem[Hagen \& Reimers (2000)]{hagen00}
Hagen, H.-J., Reimers, D.\ 2000, A\&A 357, L31

\bibitem[Hjorth et al.\ (2002)]{hjorth02}
Hjorth, J., Burud, I., Jaunsen, A. O., et al.\ 2002, \apj, 572, L11

\bibitem[Inada et al.\ (2003)]{inada03}
Inada, N., Becker, R. H., Burles, S. et al. 2003, \aj, 126, 666

\bibitem[Jakobsson et al.\ (2005)]{jakobsson04}
Jakobsson, P., Hjorth, J., Burud, I., Letawe, G., Lidman, C., Courbin, F.
2005, A\&A, 431, 103

\bibitem[Johnston et al.\ (2003)]{johnston03}
Johnston, D. E., Richards, G. T., Frieman, J. A. et al. 2003,
\aj, 126, 2281

\bibitem[Keeton et al.\ 2000]{keeton00}
Keeton, C. R., Falco, E. E., Impey, C. D., et al.\ 2000, \apj, 542, 74

\bibitem[Keeton et al.\ (2005)]{keeton05}
Keeton, C. R., Burles, S., Schechter, P. L., \& Wambsganss, J.,
{\tt astro-ph/0507521}

\bibitem[Kochanek et al.\ (1997)]{kochanek97}
Kochanek, C. S., Falco, E. E., Schild, R., et al. \ 1997, ApJ, 479, 678

\bibitem[Kochanek et al.\ (1998)]{castles}
Kochanek, C. S., Falco, E. E., Impey, C., et al.\ 1998,
{\tt cfa-www.harvard.edu/glensdata}

\bibitem[Kochanek et al.\ (2005)]{kochanek05}
Kochanek, C.S, Morgan, N.D, Falco, E.E., et al.\ 2005
{\tt astro-ph/0508070}

\bibitem[Lovell et al.\ (1998)]{lovell98}
Lovell, J. E. J., Jauncey, D. L., Reynolds, et al.\ 1998, \apjl, 508, 51

\bibitem[Lubin et al.\ (2000)]{lubin00}
Lubin, L. M., Fassnacht, C. D., Readhead, A. C. S., Blandford, R. D.,
Kundi\'c, T. 2000, \aj, 119, 451

\bibitem[Morgan et al.\ (2003a)]{morgan03a}
Morgan, N. D., Gregg, M. D., Wisotzki, L. et al.\ 2003,
\aj, 126, 696

\bibitem[Morgan et al.\ (2003b)]{morgan03b}
Morgan, N. D., Snyder, J. A., Reens, L. H. 2003,
\aj, 126, 2145

\bibitem[Morgan et al.\ (2004)]{morgan04}
Morgan, N. D., Caldwell, J. A. R., Schechter, P. L. et al.\ 2004, \aj,
127, 2617

\bibitem[Ofek \& Maoz, D. (2003)]{ofek03}
Ofek, E. O. \& Maoz, D. 2003, \apj, 594, 101

\bibitem[Oguri et al.\ 2004a]{oguri04a}
Oguri, M., Inada, N., Keeton, C. R., et al.\ 2004, \apj, 605, 78

\bibitem[Oguri et al.\ (2004b)]{oguri04b}
Oguri, M., Inada, N., Castander, F. J. et al.\ 2004, \pasj, 56, 399

\bibitem[Oscoz et al.\ (1997)]{oscoz97}
Oscoz, A., Serra-Ricart, M., Mediavilla, E., Buitrago, J., \&
Goicoechea, L. J. 1997, \apjl, 491, 7

\bibitem[Oscoz et al.\ (2001)]{oscoz01}
Oscoz, A., Alcalde, D., Serra-Ricart, M., et al.\ 2001, \apj, 552, 81

\bibitem[Patnaik et al.\ (1992)]{patnaik92}
Patnaik, A. R., Browne, I. W. A., Walsh, D., et al. \ 1992, MNRAS, 259

\bibitem[Pindor et al.\ (2004)]{pindor04} 

Pindor, B., Eisenstein, D. J., Inada, N. G., et al. \ 2004, AJ, 127,
1318
\bibitem[Refsdal (1964)]{refsdal64}
Refsdal, S. 1964, \mnras, 128, 307

\bibitem[Saha (2004)]{s04}
Saha, P. 2004, A\&A, 414, 425

\bibitem[Saha \& Williams 1997]{sw97}
Saha, P. \& Williams, L. L. R. 1997, \mnras, 292, 148-156

\bibitem[Saha \& Williams (2003)]{sw03}
Saha, P. \& Williams, L. L. R. 2003, \aj, 125, 2769

\bibitem[Saha \& Williams 2004]{sw04}
Saha, P. \& Williams, L. L. R. 2004, \aj, 127, 2604

\bibitem[Schechter et al.\ (1997)]{schechter97}
Schechter, P. L., Bailyn, C. D., Barr, R., et al.\ 1997, \apj, 475, L85

\bibitem[Sluse et al. \ (2003)]{sluse03}
 Sluse, D., Surdej, J., Claeskens, J.-F. et al. 2003, A\&A, 406, L43

\bibitem[Surdej et al. \ (1993)]{surdej93}
Surdej, J., Remy, M., Smette, A., et al. \ 1993, 153, {\sl Proc. of
the 31st Li\`ege International Astrophysical Colloquium `Gravitational
lenses in the Universe'; eds. Surdej et al.}

\bibitem[Spergel et al. 2003]{spergel03}
Spergel, D. N., Verde, L., Peiris, H. V.  et al.\ 2003, \apjs, 148, 175

\bibitem[Williams \& Saha 2000]{ws00}
Williams, L. L. R. \& Saha, P. 2000, \aj, 119, 439

\bibitem[Winn et al. \ 2002]{winn02}
Winn, J.N., Kochanek, C.S., McLeod, B.A. et al. \ 2002, ApJ 575, 103

\bibitem[Wizotski et al. (2002)]{wizotski02}
Wisotzki, L., Schechter, P. L., Bradt, H. V., Heinm\"uller, J.,
Reimers, D., A\&A, 395, 17

\bibitem[Xanthopoulos et al.\ (1998)]{xanthopoulos}
Xanthopoulos, E., Browne, I. W. A., King, L. J. 1998, \mnras, 300, 64

\end{thebibliography}
\end{document}